\documentclass[runningheads,a4paper]{llncs}

\usepackage{xcolor,colortbl}

\usepackage{graphicx}
\usepackage{amsmath}
\usepackage{amssymb}
\usepackage{mathrsfs}
\usepackage[hidelinks]{hyperref}
\usepackage{booktabs} 
\usepackage[ruled]{algorithm2e} 
\usepackage{changepage}

\begin{document}

\title{A Survey Of Methods\\ For Explaining Black Box Models} 
\titlerunning{Open The Black Box Survey}

\author{
	Riccardo Guidotti\inst{1,2}, 
	Anna Monreale\inst{1}, 
	Salvatore Ruggieri\inst{1}, 
	Franco Turini\inst{1}, 
	Dino Pedreschi\inst{1}, 
	Fosca Giannotti\inst{2}
}

\authorrunning{R. Guidotti et al.}

\institute{
	University of Pisa, \{name.surname\}@di.unipi.it
\and
	ISTI-CNR, Pisa, \{name.surname\}@isti.cnr.it
}

\maketitle

\begin{abstract}
In the last years many accurate decision support systems have been constructed as black boxes, that is as systems that hide their internal logic to the user. This lack of explanation constitutes both a practical and an ethical issue. The literature reports many approaches aimed at overcoming this crucial weakness sometimes at the cost of scarifying accuracy for interpretability. The applications in which black box decision systems can be used are various, and each approach is typically developed to provide a solution for a specific problem and, as a consequence, delineating explicitly or implicitly its own definition of interpretability and explanation. The aim of this paper is to provide a classification of the main problems addressed in the literature with respect to the notion of explanation and the type of black box system.
Given a problem definition, a black box type, and a desired explanation this survey should help the researcher to find the proposals more useful for his own work. The proposed classification of approaches to open black box models should also be useful for putting the many research open questions in perspective.

\end{abstract}


\section{Introduction}
\label{sec:introduction}

The last decade has witnessed the rise of ubiquitous opaque decision systems.
These black box systems exploit sophisticated machine learning models to predict individual information that may also be sensitive.
We can consider credit score, insurance risk, health status, as examples.
Machine learning algorithms build predictive models which are able to map user features into a class (outcome or decision) thanks to a learning phase.
This learning process is made possible by the digital traces that people leave behind them while performing everyday activities (e.g., movements, purchases, comments in social networks, etc.).
This enormous amount of data may contain human biases and prejudices.
Thus, decision models learned on them may inherit such biases, possibly leading to unfair and wrong decisions. 

The European Parliament recently adopted the \emph{General Data Protection Regulation (GDPR)}, which will become law in May 2018. 
An innovative aspect of the GDPR, which has been much debated, are the clauses on automated (algorithmic) individual decision-making, including profiling, which for the first time introduce, to some extent, a right of explanation for all individuals to obtain ``meaningful explanations of the logic involved'' when automated decision making takes place. 
Despite divergent opinions among legal scholars regarding the real scope of these clauses \cite{goodman2016eu,wachter2017right,comande2017regulating}, everybody agrees that the need for the implementation of such a principle is urgent and that it represents today a huge open scientific challenge. 
Without an enabling technology capable of explaining the logic of black boxes, the right to an explanation will remain a ``dead letter''.

By relying on sophisticated machine learning models trained on massive datasets thanks to scalable, high-performance infrastructures, we risk to create and use decision systems that we do not really understand. 
This impacts not only information on ethics, but also on safety and on industrial liability. 
Companies increasingly market services and products by embedding machine learning components, often in safety-critical industries such as self-driving cars, robotic assistants, and personalized medicine. 
Another inherent risk of these components is the possibility of inadvertently making wrong decisions, learned from artifacts or spurious correlations in the training data, such as recognizing an object in a picture by the properties of the background or lighting, due to a systematic bias in training data collection. 
How can companies trust their products without understanding and validating the underlying rationale of their machine learning components? 
Gartner predicts that ``by 2018 half of business ethics violations will occur through the improper use of Big Data analytics''. 
Explanation technologies are an immense help to companies for creating safer, more trustable products, and better managing any possible liability they may have. 
Likewise, the use of machine learning models in scientific research, for example in medicine, biology, socio-economic sciences, requires an explanation not only for trust and acceptance of results, but also for the sake of the openness of scientific discovery and the progress of research. 

As a consequence, explanation is at the heart of a responsible, open data science, across multiple industry sectors and scientific disciplines. 
Different scientific communities studied the problem of explaining machine learning decision models. 
However, each community addresses the problem from a different perspective and provides a different meaning to \textit{explanation}. 
Most of the works in the literature come from the machine learning and data mining communities. 
The first one is mostly focused on describing how black boxes work, while the second one is more interested in explaining the decisions even without understanding the details on how the opaque decision systems work in general.

Despite the fact that interpretable machine learning has been a topic for quite some time and received recently much attention, today there are many ad-hoc scattered results, and a systematic organization and classification of these methodologies is missing.
Many questions feed the papers in the literature proposing methodologies for interpreting black box systems \cite{weller2017challenges,hofman2017prediction}: 
\textit{What does it mean that a model is interpretable or transparent? What is an explanation? 
When a model or an explanation is comprehensible? 
Which is the best way to provide an explanation and which kind of model is more interpretable? 
Which are the problems requiring interpretable models/predictions? 
What kind of decision data are affected? 
Which type of data records is more comprehensible? 
How much are we willing to lose in prediction accuracy to gain any form of interpretability? }

We believe that a clear classification considering simultaneously all these aspects is needed to organize the body of knowledge about research investigating methodologies for opening and understanding the black box. 
Existing works tend to provide just a general overview of the problem \cite{lipton2016mythos} highlighting unanswered questions and problems \cite{doshi2017towards}.
On the other hand, other works focus on particular aspects like the impact of representation formats on comprehensibility \cite{huysmans2011empirical}, or the interpretability issues in term of advantages and disadvantages of selected predictive models \cite{freitas2014comprehensible}.
Consequently, after recognizing four categories of problems and a set of ways to provide an explanation, we have chosen to group the methodologies for opening and understanding black box predictors by considering simultaneously the problem they are facing, the class of solutions proposed for the explanation, the kind of data analyzed and the type of predictor explained.

The rest of the paper is organized as follows.
Firstly, in Section~\ref{sec:interpretability} we discuss what interpretability is.
Section~\ref{sec:motivations} show which are the motivations for requiring explanation for black box systems by illustrating some real cases.
In Section~\ref{sec:problem} we formalize our problem definitions used to categorize the state of the art works.
Details of the classification and crucial points distinguishing the various approaches and papers are discussed in Section~\ref{sec:classification}.
Sections \ref{sec:model_explanation}, \ref{sec:outcome_explanation}, \ref{sec:inspection} and \ref{sec:transparent_design} present the details of the  solutions proposed.
Finally, Section~\ref{sec:conclusion} summarizes the crucial aspects emerged from the analysis of the state of the art and discusses which are the open research questions and future research directions.

\section{Needs for Interpretable Models}
\label{sec:motivations}
\textit{Which are the real problems requiring interpretable models and explainable predictions?}
In this section, we briefly report some cases showing how and why black boxes can be dangerous. 
Indeed, delegating decisions to black boxes without the possibility of an interpretation may be critical, can create discrimination and trust issues.

Training a classifier on historical datasets, reporting human decisions, could lead to the discovery of endemic preconceptions 
\cite{pedreshi2008discrimination}.
Moreover, since these rules can be deeply concealed within the trained classifier, we risk to consider, maybe unconsciously, such practices and prejudices as general rules. 
We are warned about a growing ``black box society'' \cite{pasquale2015black}, governed by ``secret algorithms protected by industrial secrecy, legal protections, obfuscation, so that intentional or unintentional discrimination becomes invisible and mitigation becomes impossible.''

Automated discrimination is not new and is not necessarily due to ``black box'' models.
A computer program for screening job applicants were used during the 1970s and 1980s in St. George's Hospital Medical School (London).  
The program used information from applicants' forms, without any reference to ethnicity. 
However, the program was found to unfairly discriminate against ethnic minorities and women by inferring this information from surnames and place of birth, and lowering their chances of being selected for interview \cite{lowry1988blot}. 
 
More recently, the journalists of \emph{propublica.org} have shown that the COMPAS score, a predictive model for the ``risk of crime recidivism'' (proprietary secret of Northpointe), has a strong ethnic bias.
Indeed, according to this score, a black who did not re-offend were classified as high risk twice as much as whites who did not re-offend, and white repeat offenders were classified as low risk twice as much as black repeat offenders\footnote{\url{http://www.propublica.org/article/machine-bias-risk-assessments-in-criminal-sentencing}}.

Similarly, a study at Princeton \cite{caliskan2016semantics} shows how text and web corpora contain human biases: names that are associated with black people are found to be significantly more associated with unpleasant than with pleasant terms, compared to names associated with whites.
As a consequence, the models learned on such text data for opinion or sentiment mining have a possibility of inheriting the prejudices reflected in the data. 

Another example is related to Amazon.com.
In 2016, the software used to determine the areas of the US to which Amazon would offer free same-day delivery, unintentionally restricted minority neighborhoods from participating in the program (often when every surrounding neighborhood was allowed)\footnote{\url{http://www.techinsider.io/how-algorithms-can-be-racist-2016-4}}.

With respect to credit bureaus, it is shown in \cite{carter2006credit} that banks providing credit scoring for millions of individuals, are often discordant: in a study of $500,000$ records, $29\%$ of consumers received credit scores that differed by at least fifty points among three major US banks (Experian, TransUnion, and Equifax).
Such a difference might mean tens of thousands of dollars over the life of a mortgage. 
So much variability implies that the three scoring systems either have a very different and undisclosed bias, or are highly arbitrary.

As example of bias we can consider \cite{freitas2014comprehensible} and \cite{ribeiro2016should}.
In these works, the authors show how accurate black box classifiers may result from an accidental artifact in the training data.
In \cite{freitas2014comprehensible} the military trained a classifier to recognize enemy tanks from friendly tanks.
The classifier resulted in a high accuracy on the test set, but when it was used in the field had very poor performance.
Later was discovered that enemy photos were taken on overcast days, while friendly photos on sunny days.
Similarly, in \cite{ribeiro2016should} is shown that a classifier trained to recognize wolves and husky dogs were basing its predictions to classify a wolf solely on the presence of snow in the background.

Nowadays, \emph{Deep Neural Networks (DNNs)} have been reaching very good performances on different pattern-recognition tasks such as visual and text classification which are easily performed by humans: e.g., saying that a tomato is displaced in a picture or that a text is about a certain topic.
Thus, what differences remain between DNNs and humans?
Despite the excellent performance of DNNs it seems to be a lot.
In \cite{szegedy2013intriguing} it is shown the alteration of an image (e.g. of a tomato) such that the change is undetectable for humans can lead a DNN to tag the image as something else (e.g., labeling a tomato as a dog). 
In \cite{nguyen2015deep} a related result is shown.
It is easy to produce images that DNNs believe to be recognizable with 99.99\% confidence, but which are completely unrecognizable to humans (e.g., labeling white static noise as a tomato). 
Similarly in \cite{koh2017understanding} visually-indistinguishable training-set are created using DNNs and linear models.
With respect to text, in \cite{liang2017deep} effective methods to attack DNN text classifiers are presented. 
Experiments show that the perturbations introduced in the text are difficult to be perceived by a human but are still able to fool a state-of-the-art DNN to misclassify a text as any desirable class. 
These results show interesting differences between humans and DNNs, and raise reasonable doubts about trusting such black boxes.
In \cite{zhang2016understanding} it is shown how conventional regularization and small generalization error fail to explain why DNNs generalize well in practice. 
Specifically, they prove that established state-of-the-art CNN trained for image classification easily fits a random labeling of the training data. 
This phenomenon occurs even if the true images are replaced by unstructured random noise.

\section{Interpretable, Explainable and Comprehensible Models}
\label{sec:interpretability}
Before presenting the classification of the problems addressed in the literature with respect to black box predictors, and the corresponding solutions and models categorization, it is crucial to understand what \emph{interpretability} is.
Thus, in this section, we discuss what an interpretable model is, and we analyze the various dimensions of interpretability as well as the desiderata for an interpretable model.
Moreover, we also discuss the meaning of words like \textit{interpretability}, \textit{explainability} and \textit{comprehensibility} which are largely used in the literature.

To \emph{interpret} means to give or provide the meaning or to explain and present in understandable terms some concept\footnote{\url{https://www.merriam-webster.com/}}.
Therefore, in data mining and machine learning, \emph{interpretability} is defined as the ability to explain or to provide the meaning in understandable terms to a human \cite{doshi2017towards}.
These definitions assume implicitly that the concepts expressed in the understandable terms composing an explanation are self-contained and do not need further explanations.
Essentially, an explanation is an ``interface'' between humans and a decision maker that is at the same time both an accurate proxy of the decision maker and comprehensible to humans.

As shown in the previous section another significant aspect to mention about interpretability is the reason why a system, a service or a method should be interpretable. 
On the other hand, an explanation could be not required if there are no decisions that have to be made on the outcome of the prediction.
For example, if we want to know if an image contains a cat or not and this information is not required to take any sort of crucial decision, or there are no consequences for unacceptable results, then we do not need an interpretable model, and we can accept any black box.

\subsection{Dimensions of Interpretability}
In the analysis of the interpretability of predictive models, we can identify a set of dimensions to be taken into consideration, and that characterize the interpretability of the model \cite{doshi2017towards}.

\emph{Global and Local Interpretability}: A model may be completely interpretable, i.e., we are able to understand the whole logic of a model and follow the entire reasoning leading to all the different possible outcomes. 
In this case, we are speaking about \textit{global} interpretability.  
Instead, we indicate with \textit{local} interpretability the situation in which it is possible to understand only the reasons for a specific decision: only the single prediction/decision is interpretable.

\emph{Time Limitation}: An important aspect is the time that the user is available or is allowed to spend on understanding an explanation. 
The user time availability is strictly related to the scenario where the predictive model has to be used. 
Therefore, in some contexts where the user needs to quickly take the decision (e.g., a disaster is imminent), it is preferable to have an explanation simple to understand. 
While in contexts where the decision time is not a constraint (e.g., during a procedure to release a loan) one might prefer a more complex and exhaustive explanation.

\emph{Nature of User Expertise}: 
Users of a predictive model may have different background knowledge and experience in the task: decision-makers, scientists, compliance and safety engineers, data scientists, etc.
Knowing the user experience in the task is a key aspect of the perception of interpretability of a model. 
Domain experts may prefer a larger and more sophisticated model over a smaller and sometimes more opaque one.

The works reviewed in the literature only implicitly specify if their proposal is global or local.
Just a few of them take into account the nature of user expertise \cite{gibbons2013cad,ribeiro2016should,schetinin2007confident}, and no one provides real experiments about the time required to understand an explanation.
Instead, some of the works consider the ``complexity'' of an explanation through an approximation.
For example, they define the model complexity as the model's size (e.g. tree depth, number of rules, number of conjunctive terms) \cite{deng2014interpreting,hara2016making,johansson2004accuracy,ribeiro2016should}.
In the following, we further discuss issues related to the complexity of an explanation.

\subsection{Desiderata of an Interpretable Model}
An interpretable model is required to provide an explanation.
Thus, to realize an interpretable model it is necessary to take into account the following list of desiderata which are mentioned by a set of papers in the state of art \cite{andrews1995survey,doshi2017towards,freitas2014comprehensible,johansson2004truth}:

\begin{itemize}

    \item \emph{Interpretability}: to which extent the model and or the prediction are human understandable. 
    The most addressed discussion is related to how the interpretability can be measured. 
    In \cite{freitas2014comprehensible} a component for measuring the interpretability is the \emph{complexity} of the predictive model in terms of the model size.
    According to the literature, we refer to interpretability also with the name \textit{comprehensibility}.
    
    \item \emph{Accuracy}: to which extent the model accurately predict unseen instances. 
    The accuracy of a model can be measured using various evaluation measures like the accuracy score, the F1-score \cite{tan2006introduction}, etc.
    Producing an interpretable model maintaining competitive levels of accuracy is the most common target among the papers in the literature.
    
    \item \emph{Fidelity}: to which extent the model is able to accurately \textit{imitate} a black-box predictor. 
    The fidelity captures how much is good an interpretable model in the mimic of the behavior of a black-box.
    Similarly to the accuracy, the fidelity is measured in terms of accuracy score, F1-score, etc. but with respect to the outcome of the black box which is considered as an oracle.
    
\end{itemize}

Moreover, besides these features strictly related to interpretability, yet according to \cite{andrews1995survey,doshi2017towards,freitas2014comprehensible,johansson2004truth} a data mining and machine learning model should have other important desiderata. 
Some of these desiderata are related to ethical aspects such as \textit{fairness} and \textit{privacy}. 
The first principle requires that the model guarantees the protection of groups against (implicit or explicit) discrimination \cite{romei2014multidisciplinary}; while the second one requires that the model does not reveal sensitive information about people \cite{aldeen2015comprehensive}. 
The level of interpretability of a model together with the standards of privacy and non-discrimination which are guaranteed may impact on how much human users trust that model. 
The degree of trust on a model increases if the model is built by respecting constraints of \textit{monotonicity} given by the users 
\cite{martens2011performance,pazzani2001acceptance,verbeke2011building}. 
A predictor respecting the \textit{monotonicity} principle is, for example, a predictor where the increase of the values of a numerical attribute 
tends to either increase or decrease in a monotonic way the probability of a record of being member of a class \cite{freitas2014comprehensible}.
Another property that influences the trust level of a model is \textit{usability}: people tend to trust more models providing information that assist them to accomplish a task with awareness.
In this line, an interactive and queryable explanation results to be more usable than a textual and fixed explanation.

Data mining and machine learning models should have other important desiderata such as \textit{reliability}, \textit{robustness}, \textit{causality}, \emph{scalability} and \emph{generality}. 
This means that a model should have the ability to maintain certain levels of performance independently from the parameters or from the input data (\textit{reliability}/\textit{robustness}) and that controlled changes in the input due to a perturbation affect the model behavior (\textit{causality}). 
Moreover, since we are in the Big Data era, it is opportune to have models able to \textit{scale} to large input data with large input spaces. 
Finally, since often in different application scenarios one might use the same model with different data, it is preferable to have portable models that do not require special training regimes or restrictions (\emph{generality}).

\subsection{Recognized Interpretable Models}
\label{sec:recognized_interpretable_models}
In the state of the art a small set of existing interpretable models is recognized: \emph{decision tree}, \emph{rules}, \emph{linear models} \cite{freitas2014comprehensible,huysmans2011empirical,ribeiro2016should}. 
These models are considered easily understandable and interpretable for humans.

\begin{figure}[!tb]
\centering
	\includegraphics[trim = 0mm 0mm 0mm 0mm, clip, width=0.4\linewidth]{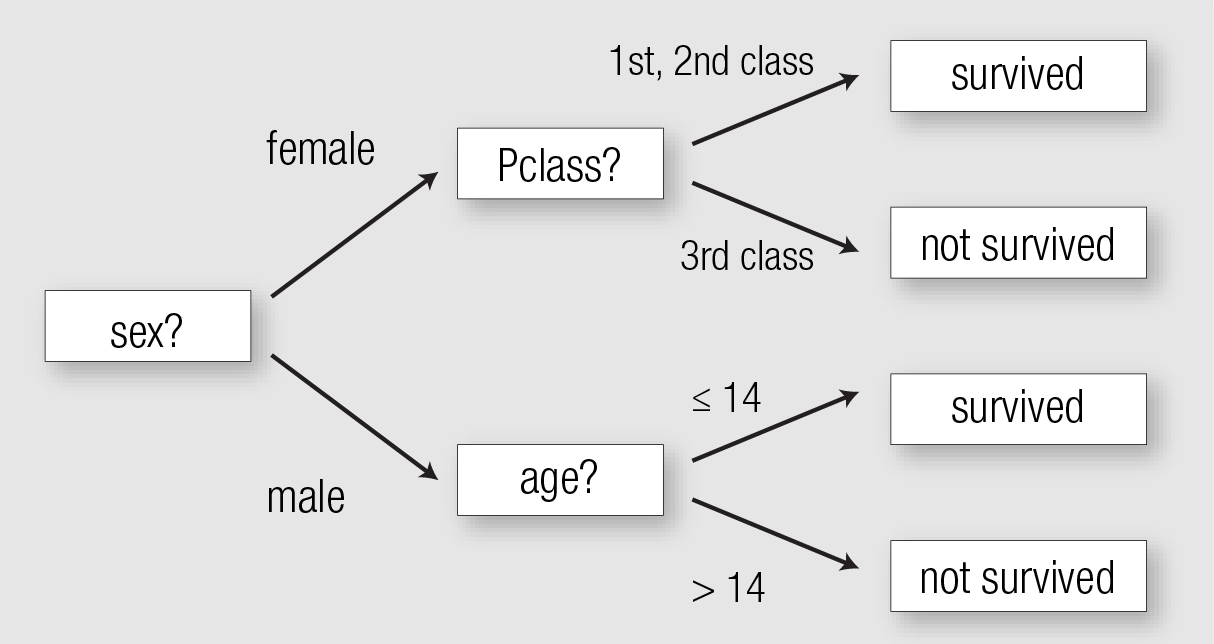}
	\caption{Example of decision tree.}
	\label{fig:decisiontree}
\end{figure}

A decision system based on a \emph{decision tree} exploits a graph structured like a tree and composed of internal nodes representing tests on features or attributes (e.g., whether a variable has a value lower than, equals to or grater than a threshold, see Figure~\ref{fig:decisiontree}), and leaf nodes representing a class label. Each branch represents a possible outcome \cite{quinlan1993c4}.
The paths from the root to the leaves represent the classification rules.
Indeed, a decision tree can be linearized into a set of decision rules with the \emph{if-then} form  \cite{quinlan1987generating,quinlan1987simplifying,frank1998generating}:
$$\mbox{if } condition_1 \wedge condition_2 \wedge condition_3 \mbox{ then } outcome.$$
Here, the outcome corresponds to the class label of a leaf node while the conjunctions of conditions in the if clause correspond to the different conditions in the path from the root to that leaf node.

More generally, a \emph{decision rule} is a function which maps an observation to an appropriate action.
Decision rules can be extracted by generating the so-called \textit{classification rules}, i.e., association rules that in the consequence have the class label \cite{agrawal1994fast}. 
The most common rules are \emph{if-then rules} where the if clause is a combination of conditions on the input variables. In particular, it may be formed by conjunctions, negations and disjunctions. However, methods for rule extraction typically take into consideration only rules with conjunctions.
Other types of rules are: 
\emph{m-of-n rules} where given a set of $n$ conditions if $m$ of them are verified then the consequence of the rule is considered true \cite{murphy1991id2};
\emph{list of rules} where given an ordered set of rules is considered true the consequent of the first rule which is verified \cite{yin2003cpar};
\emph{falling rule lists} consists of a list of if-then rules ordered with respect to the probability of a specific outcome and the order identifies the example to be classified by that rule \cite{wang2015falling};
\emph{decision sets} where an unordered set of classification rules is provided such that the rules are not connected by else statements, but each rule is an independent classifier that can assign its label without regard for any other rules \cite{lakkaraju2016interpretable}.

The interpretation of rules and decision trees is different with respect to different aspects \cite{freitas2014comprehensible}.
Decision trees are widely adopted for their graphical representation, while rules have a textual representation.  
The main difference is that textual representation does not provide immediately information about the more relevant attributes of a rule.
On the other hand, the hierarchical position of the features in a tree gives this kind of clue. 

Attributes' relative importance could be added to rules by means of positional information.
Specifically, rule conditions are shown by following the order in which the rule extraction algorithm added them to the rule.
Even though the representation of rules causes some difficulties in understanding the whole model, it enables the study of single rules representing partial parts of the whole knowledge (``local patterns'') which are composable. 
Also in a decision tree, the analysis of each path separately from the leaf node to the root, enables users to focus on such local patterns.
However, if the tree is very deep in this case it is a much more complex task.
A further crucial difference between rules and decision trees is that in a decision tree each record is classified by only one leaf node, i.e., the class predicted are represented in a mutually exclusive and exhaustive way by the set of leaves and their paths to the root node. 
On the other hand, a certain record can satisfy the antecedent of rules having as consequent a different class for that record.
Indeed, rule based classifiers have the disadvantage of requiring an additional approach for resolving such situations of conflicting outcome \cite{wettschereck1997review}. 
Many rule based classifiers deal with this issue by returning an ordered rule list, instead of an unordered rule set.
In this way it is returned the outcome corresponding to the first rule matching the test record and ignoring the other rules in the list.
We notice that ordered rule lists may be harder to interpret than classical rules.
In fact, in this model a given rule cannot be considered independently from the precedent rules in the list 
\cite{wettschereck1997review}.
Another widely used approach consists in considering the top-k rules satisfying the test record where the ordering is given by a certain weight (e.g. accuracy, Laplace accuracy, etc.).
Then, the outcome of the rules with the average highest weight among the top-k is returned as predicted class \cite{yin2003cpar}.

\begin{figure}[!tb]
\centering
	\includegraphics[trim = 0mm 0mm 0mm 0mm, clip, width=0.4\linewidth]{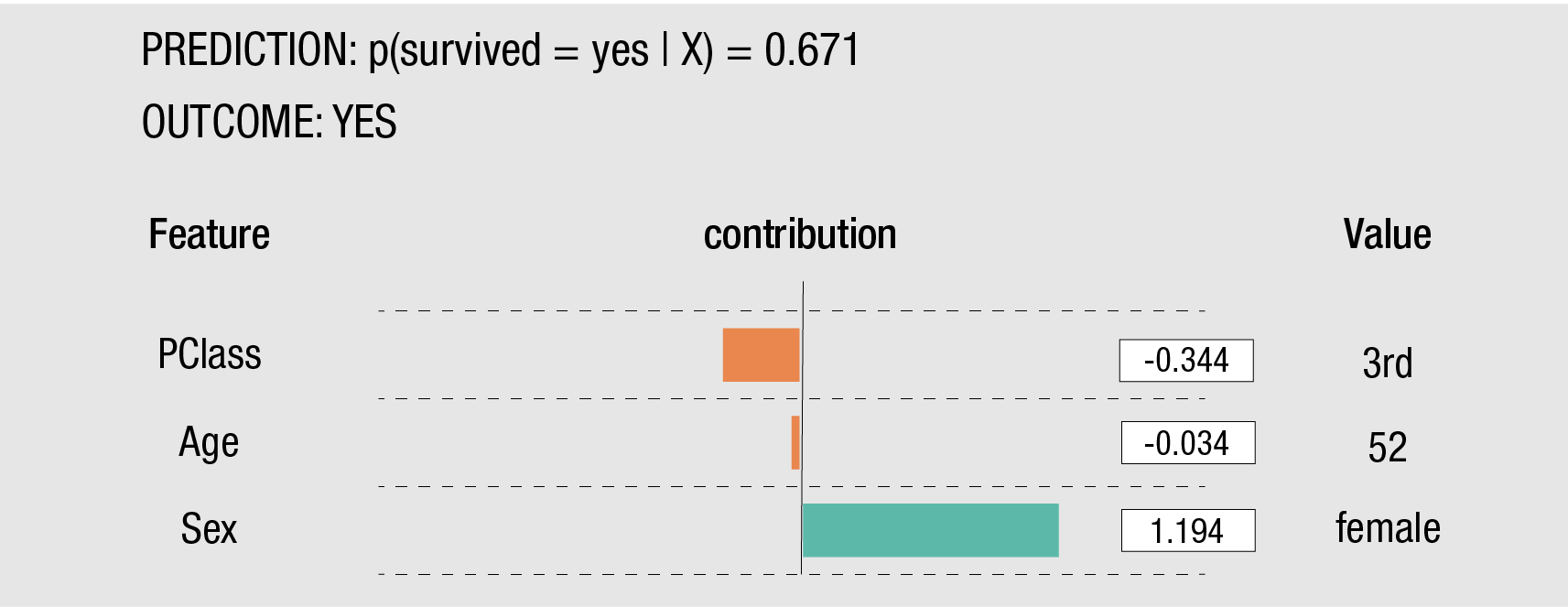}
	\caption{Example of feature importance for a linear model.}
	\label{fig:linearmodel}
\end{figure}

Finally, explanations can also be provided through linear models \cite{kononenko2010efficient,ribeiro2016should}.
This can be done by considering and visualizing both the sign and the magnitude of the contribution of the attributes for a given prediction (see Figure \ref{fig:linearmodel}). 
If the contribution of an attribute-value is positive, then it contributes by increasing the model's output. Instead, if the sign is negative then the attribute-value decreases the output of the model. 
If an attribute-value has an higher contribution than another, then it means that it has an higher influence on the prediction of the model. 
The produced contributions summarize the performance of the model, thus the difference between the predictions of the model and expected predictions, providing the opportunity of quantifying the changes of the model prediction for each test record. In particular, it is possible to identify the attributes leading to this change and for each attribute how much it contributed to the change.

As last remark we point out that in general, when an explanation for a prediction is provided, it is often useful to analyze besides the explanation (satisfied rules, branch of the tree, set of weights, etc.), also instances which are exceptions with respect to the ``boundaries'' provided by the explanation, or with very few differences with respect to the prototypes returned as explanation. 
For example, instances covered by the rule body but with an outcome label different from the class of the outcome predicted. 
Even though this sort of \emph{exception analysis} is hardly performed, it can be more informative than the direct explanation, and it can also provide clues about the application domain \cite{pappa2005predicting}.

\subsection{Explanations and Interpretable Models Complexity}
In the literature, very little space is dedicated to a crucial aspect: the model complexity.
The evaluation of the model complexity is generally tied to the model comprehensibility, and this is a very hard task to address.
As a consequence, this evaluation is generally estimated with a rough approximation related to the size of the model. 
Moreover, complexity is often used as an opposed term to interpretability.

In \cite{hara2016making} the complexity is identified by the number of regions, i.e., the parts of the model, for which the boundaries are defined.
In \cite{ribeiro2016should} as complexity for linear models is adopted the number of non-zero weights, while for decision trees the depth of the tree.
In \cite{deng2014interpreting} the complexity of a rule (and thus of an explanation) is measured by the length of the rule condition, defined as the number of attribute-value pairs in the condition. 
Given two rules with similar frequency and accuracy, the rule with a smaller length may be preferred as it is more interpretable. 
Similarly, in case of lists of rules the complexity is typically measured considering the total number of attribute-value pairs in the whole set of rules.
However, this could be a suitable way for measuring the model complexity, since in an ordered rule list different test records need distinct numbers of rules to be evaluated \cite{freitas2014comprehensible}. 
In this kind of model, a more honest measure could be the average number of conditions evaluated to classify a set of test records \cite{otero2013improving}.
However, this is more a ``measure of the explanation'' of a list of rules. 

Differently from the not flexible representation of decision tree where the prediction of a single record is mutually exhaustive and exclusive, rules characterization contains only significant clauses.
As a consequence, an optimal set of rules does not contain any duplicated information, given the fact that an outcome label can appear only one time in the consequent of a set of rules, while in a decision tree it typically comes out more than once.
Moreover, rules do not capture insignificant clauses, while decision trees can also have insignificant branches. 
This happens because rule based classifier generally select one \emph{attribute-value} while expanding a rule, whereas decision tree algorithms usually select one \emph{attribute} while expanding the tree \cite{freitas2014comprehensible}. 
Considering these aspects to estimate the complexity is very difficult. 
Consequently, even though a model equivalence exists, the estimation of the fact that a different representation for the same model (or explanation) is more complex than another when using decision trees or rules can be very subjective with respect to the interpreter.

\subsection{Interpretable Data for Interpretable Models}
The \textit{types of data} used for classification may have diverse nature. 
Different types of data present a different level of interpretability for a human. 
The most understandable data format for humans is the \emph{table} \cite{huysmans2011empirical}. 
Since matrices and vectors are the typical data representation used by the vast majority of data mining and machine learning techniques, tables are also easily managed by these algorithms without requiring specific transformations.

Other forms of data which are very common in human daily life are \emph{images} and \emph{texts}. 
They are perhaps for human brain even more easily understandable than tables. 
On the other hand, the processing of these data for predictive models requires their transformation into vectors that make them easier to process by algorithms but less interpretable for humans.
Indeed, on images and texts, the state of art techniques typically apply predictive models based on super vector machine, neural networks or deep neural networks that are usually hard to be interpreted.
As a consequence, certain recognized interpretable models cannot be directly employed for this type of data in order to obtain an interpretable model or a human understandable explanation.
Transformations using equivalences, approximations or heuristics are required in such a way that images and texts can be employed by prediction systems and used for providing the interpretation of the model and/or the prediction at the same time. 

Finally, there exist other forms of data such as sequence data, spatio-temporal data and complex network data that may be used by data mining and machine learning algorithms.
However in the literature, to the best of our knowledge, there is no work addressing the interpretability of models for data different from images, texts, and tabular data.

\section{Open The Black Box Problems}
\label{sec:problem}
An accurate analysis and review of the literature lead to the identification of different categories of problems.

At a very high level, we can distinguish between \emph{reverse engineering} and \emph{design} of explanations.
In the first case, given the decision records produced by a black box decision maker the problem consists in reconstructing an explanation for it. 
The original dataset upon which the black box is trained is generally not known in real life.
Details about reverse engineering approaches are discussed at the end of this section.
On the other hand, it is used and exploited to build the explanations by most of the works presented in this survey.
In the second case, given a dataset of training decision records the task consists in developing an interpretable predictor model together with its explanations.

Through a deep analysis of the state of the art we are able to further refine the first category obtaining three different problems.
We name them \textit{black box model explanation problem}, \textit{black box outcome explanation problem}, and \textit{black box inspection problem}.
We name the second category \textit{transparent box design problem}.
All these problems can be formalized as specific cases of the general classification problems with the common target of providing an interpretable and accurate predictive model.
Details of the formalization are provided in the following sections.
Other important variants are generally not treated in the literature making the problem of discovering an explanation increasingly difficult: 
\emph{(i)} Is it allowed to query the black box at will to obtain new decision examples, or only a fixed dataset of decision records is available? 
\emph{(ii)} Is the complete set of features used by the decision model known, or instead only part of these features is known?
In this survey we do not address these issues as in the literature there is not sufficient material.

\subsection{Problem Formulation}
\begin{figure}[!tb]
\centering
	\includegraphics[trim = 0mm 0mm 0mm 0mm, clip, width=0.48\linewidth]{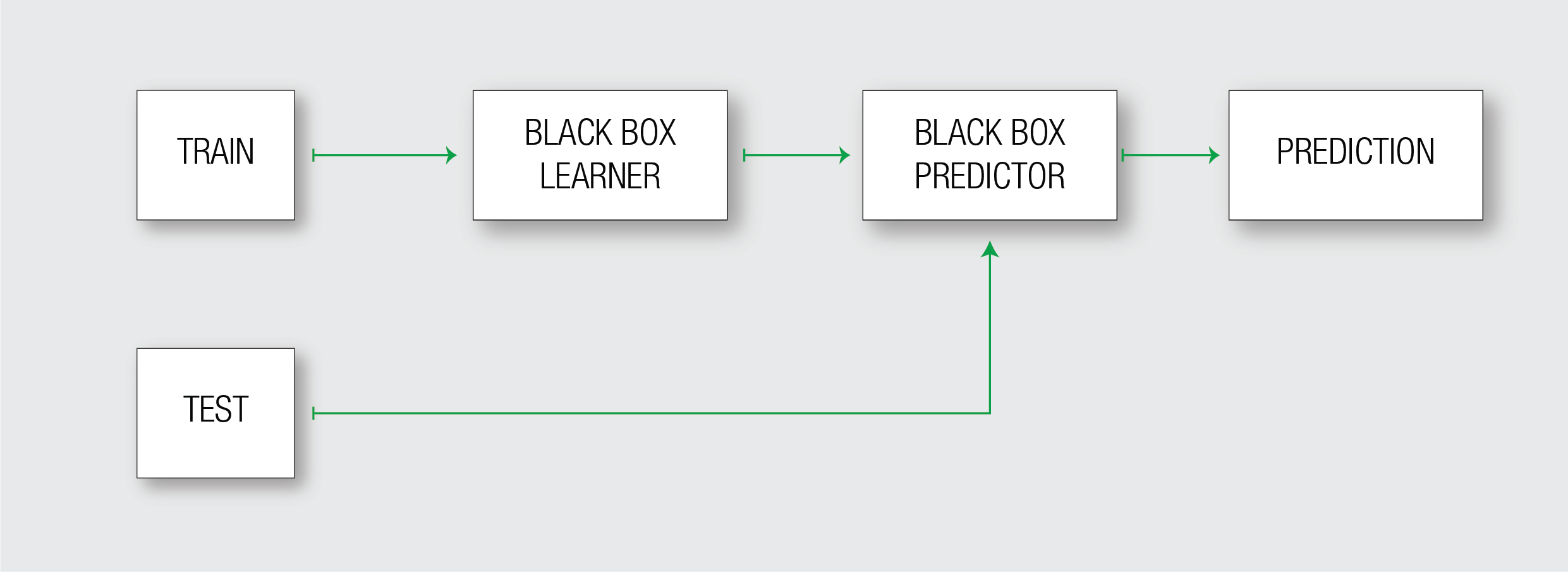}
	\caption{Classification Problem.}
	\label{fig:problem}
\end{figure}

In the following, we generalize the classification problem (see Figure~\ref{fig:problem}).

A \emph{predictor}, also named model or classifier, is a function $b:\mathcal{X}^m \rightarrow \mathcal{Y}$ where $\mathcal{X}^m$ is the \emph{feature space} with $m$ corresponding to the number of features, and $\mathcal{Y}$ is the \emph{target space}. 
The feature space $\mathcal{X}$ can correspond to any basic data type like the set of integers $\mathcal{X} = \mathbb{I}^m$, 
reals $\mathcal{X} = \mathbb{R}^m$, 
booleans $\mathcal{X} = \{0,1\}^m$,
and strings $\mathcal{X} = S^m$, where $S=\Sigma^*$ and $\Sigma = \{ a, b, c, \dots, \}$ is the alphabet (a finite non-empty set of symbols). 
The feature space $\mathcal{X}$ can also be a complex data type composed of different basic data type. 
For example, $\mathcal{X} = \mathbb{I} \times \mathbb{R}^2 \times S$ contains an integer feature, two real features and a string feature.
On the other hand, the target space $\mathcal{Y}$ (with dimensionality equals to one) contains the different labels (classes or outcomes) and identifies a semantic concept where $\mathcal{Y}$ can be a set of booleans, integers or strings.

A predictor $b$ is the output of a \emph{learner} function $\mathscr{L}_b$ such that $\mathscr{L}_b : (\mathcal{X}^{n \times m} \times \mathcal{Y}^n) \rightarrow (\mathcal{X}^m \rightarrow \mathcal{Y})$.
The learner $\mathscr{L}_b$ takes as input a dataset $D = \{ X, Y \}$ with $n$ samples where $X \in \mathcal{X}^{n \times m}$ and $Y \in \mathcal{Y}^n$ and returns the predictor $b$. 
Given a data record in the feature space $x \in \mathcal{X}^m $, the predictor $b$ can be employed to predict the target value \mbox{$\hat y$, i.e., $b(x) = \hat{y}$.}

Typically, in supervised learning \cite{tan2006introduction}, a training dataset $D_{\mathit{train}}$ is used for training the learner $\mathscr{L}_b(D_{\mathit{train}})$ which builds the predictor $b$, and a test dataset $D_{\mathit{test}}$ is used for evaluating the performance of $b$.
Given $D_{\mathit{test}} = \{X, Y\}$, the evaluation is performed by observing for each couple of data record and target value $(x, y) \in D_{\mathit{test}}$ the number of correspondences between $y$ and $b(x) = \hat{y}$.

In the following we indicate with $b$ a black box predictor belonging to the set of \textit{uninterpretable} data mining and machine learning models. 
According to Section \ref{sec:interpretability}, $b$ is a black box because the reasoning behind the function is not understandable by humans and the outcome returned does not provide any clue for its choice. 
In real-world applications, $b$ is an opaque classifier resulting from a learning $\mathcal{L}_b$.
Similarly, we indicate with $c$ a comprehensible predictor for which is available a global or a local explanation.

The performance of the comprehensible predictor $c$ is generally evaluated by two measures. 
The \textit{accuracy} is used to evaluate how good are the performance of both the black box predictor $b$ and the comprehensible predictor $c$. 
The \textit{fidelity} is employed to evaluate how good is the comprehensible predictor $c$ in mimicking the black box predictor $b$.
Indeed, given a data set $D=\{X,Y\}$ we can apply to each record $x \in X$ both the predictors: \emph{(i)} for the black box $b$ we get the set of predictions $\hat{Y} = \bigcup_{x\in X} b(x)$, while \emph{(ii)} for the comprehensible predictor $c$ we get the set of predictions $\bar{Y} = \bigcup_{x\in X} c(x)$.

Thus, we can evaluate the accuracy of the black box $b$ and of the comprehensible predictor $c$ by comparing the real target values $Y$ against the predicted target values $\hat{Y}$, and $\bar{Y}$ with $\mathit{accuracy}(\hat{Y}, Y)$ and $\mathit{accuracy}(\bar{Y}, Y)$, respectively.
Moreover, we can evaluate the behavior of the predictor $c$ with respect to $b$ evaluating the fidelity of $c$ by means of the function $\mathit{fidelity}(\hat{Y}, \bar{Y})$.
Note that the $\mathit{fidelity}$ score can be calculated by applying the same calculus of the $\mathit{accuracy}$ function where as target value is used the prediction $\bar{Y}$ of the black box $b$ instead of the real values $Y$.

\subsection*{Black Box Model Explanation} 
\label{sec:model_explanation_problem}
\begin{figure}[!tb]
\centering
	\includegraphics[trim = 0mm 0mm 0mm 0mm, clip, width=0.48\linewidth]{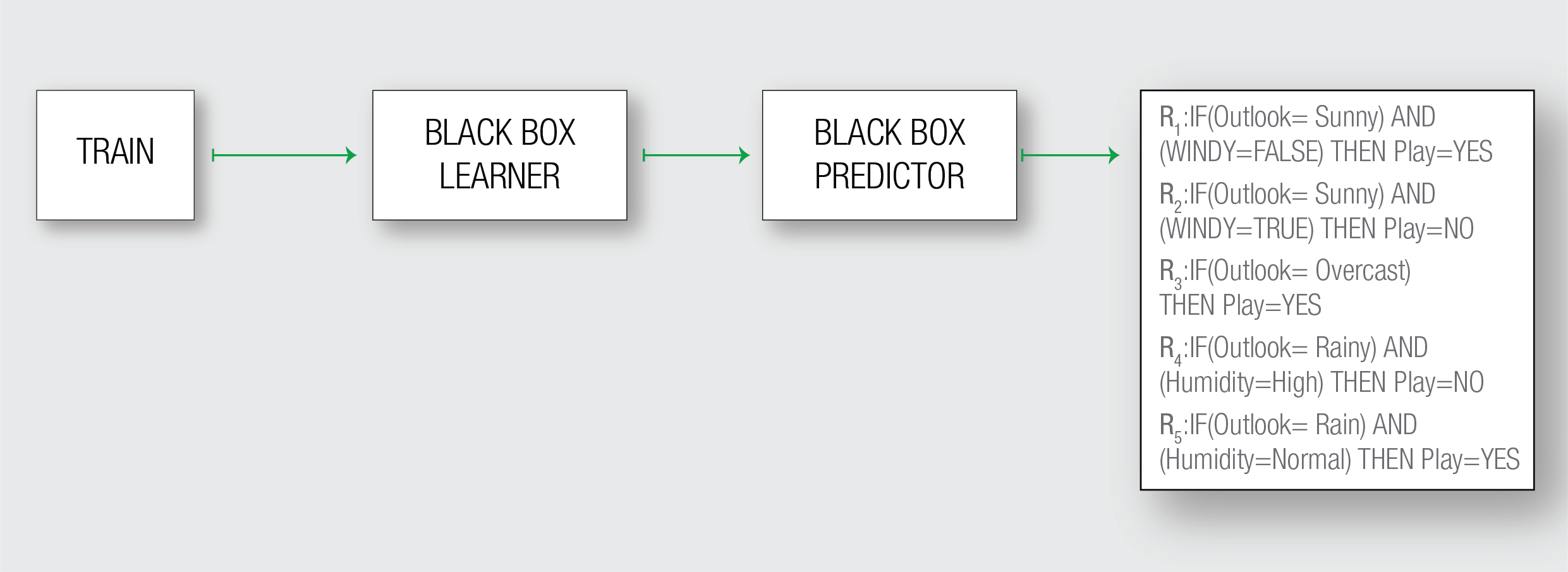}
		\caption{Black Box Model Explanation Problem.}
	\label{fig:prob_model_expl}
\end{figure}

Given a black box model solving a classification problem, \emph{the black box explanation problem} consists in providing an interpretable and transparent model which is able to mimic the behavior of the black box and which is also understandable by humans (see Figure~\ref{fig:prob_model_expl}). 
In other words, the interpretable model approximating the black box must be globally interpretable.
As consequence, we define the black box model explanation problem as follows:

\begin{definition}[Black Box Model Explanation]
Given a black box predictor $b$ and a dataset $D = \{ X, Y \}$, the \emph{black box model explanation problem} consists in finding a function $f: (\mathcal{X}^m \rightarrow \mathcal{Y}) \times (\mathcal{X}^{n \times m} \times \mathcal{Y}^n) \rightarrow (\mathcal{X}^m \rightarrow \mathcal{Y})$ which takes as input a black box $b$ and a dataset $D$, and returns a comprehensible global predictor $c_g$, i.e., $f(b, D) = c_g$, such that $c_g$ is able to mimic the behavior of $b$, and exists a \textit{global explanator} function $\varepsilon_g: (\mathcal{X}^m \rightarrow \mathcal{Y}) \rightarrow \mathcal{E}$ that can derive from $c_g$ a set of explanations $E \in \mathcal{E}$ modeling in a human understandable way the logic behind 
$c_g$, i.e., $\varepsilon_g(c_g) = E$.
\end{definition}

A large set of the papers reviewed in this survey describe various designs for the function $f$ to solve the black box explanation problem.
The set of explanations $E$ can be modeled for example by a decision tree or by a set of rules \cite{huysmans2011empirical}, while the comprehensible global predictor $c_g$ is the predictor returning as global explanation $\varepsilon_g$ the decision tree or the set of rules.

\subsection*{Black Box Outcome Explanation} 
\label{sec:outcome_explanation_problem}
Given a black box model solving a classification problem, the \textit{black box outcome explanation problem} consists in providing an interpretable outcome, that is a method for providing an explanation for the outcome of the black box. 
In other words, the interpretable model must return the prediction together with an explanation about the reasons for that prediction, i.e., the prediction  is only locally interpretable.
It is not required to explain the whole logic behind the black box but only the reasons for the choice of a particular instance.
Consequently, we define the black box outcome explanation problem as:

\begin{definition}[Black Box Outcome Explanation]
Given a black box predictor $b$ and a dataset $D = \{ X, Y \}$, the \emph{black box outcome explanation problem} consists in finding a function $f: (\mathcal{X}^m \rightarrow \mathcal{Y}) \times (\mathcal{X}^{n \times m} \times \mathcal{Y}^n) \rightarrow (\mathcal{X}^m \rightarrow \mathcal{Y})$ which takes as input a black box $b$ and a dataset $D$, and returns a comprehensible local predictor $c_l$, i.e., $f(b, D) = c_l$, such that $c_l$ is able to mimic the behavior of $b$, and exists a \textit{local explanator} function $\varepsilon_l: ((\mathcal{X}^m \rightarrow \mathcal{Y}) \times (\mathcal{X}^m \rightarrow \mathcal{Y}) \times \mathcal{X}^m) \rightarrow \mathcal{E}$ which takes as input the black box $b$, the comprehensible local predictor $c_l$, and a data record $x$ with features in $\mathcal{X}^m$, and returns a human understandable \textit{explanation} $e \in \mathcal{E}$ for the data record $x$, i.e., $\varepsilon_l(b, c_l, x) = e$.
\end{definition}

We report in this survey recent works describing very diversified approaches to implement function $f$, overcoming the limitations of explaining the whole model (illustrated in Section \ref{sec:model_explanation}). 
As an example, in this view of the problem, we can consider that the explanation $e_l$ may be either a path of a decision tree or an association rule \cite{freitas2014comprehensible}.

\begin{figure}[!tb]
\centering
	\includegraphics[trim = 0mm 0mm 0mm 0mm, clip, width=0.48\linewidth]{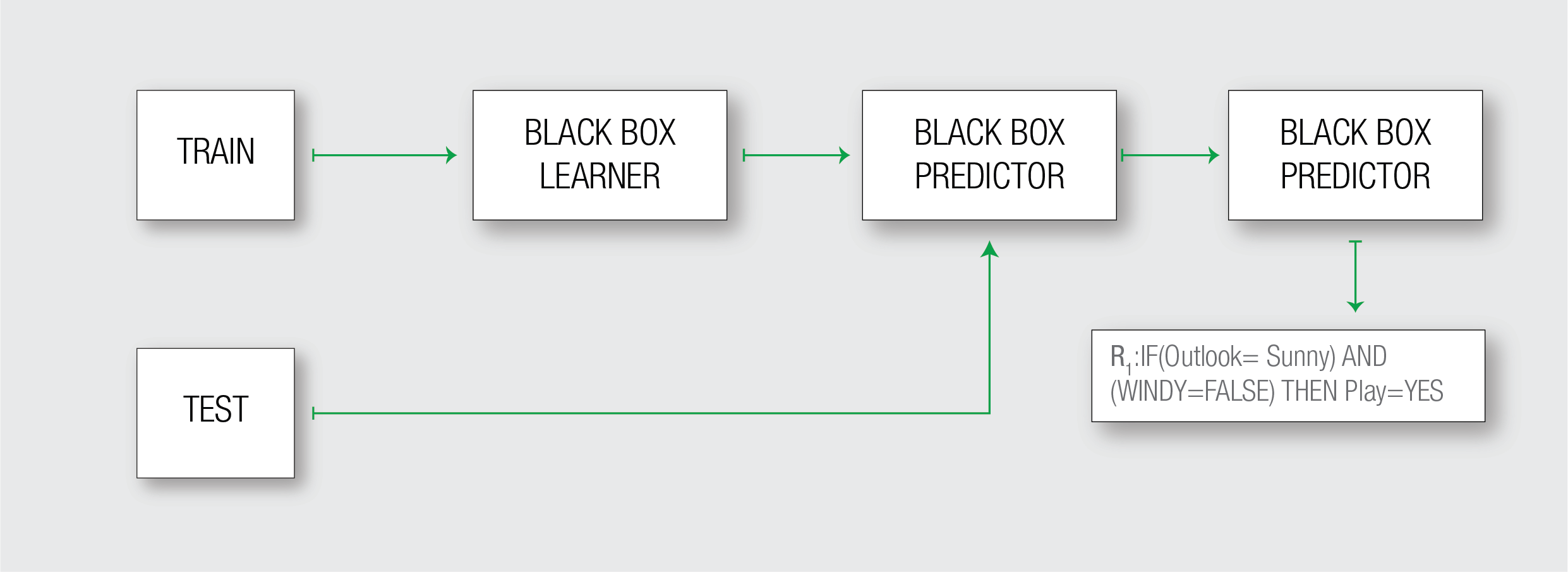}
		\caption{Black Box Outcome Explanation Problem.}
	\label{fig:prob_outcome_expl}
\end{figure}

\subsection*{Black Box Inspection Problem}
\label{sec:inspection_problem}
Given a black box model solving a classification problem, the \textit{black box inspection problem} consists in providing a representation (visual or textual) for understanding either how the black box model works or why the black box returns certain predictions more likely than others.

\begin{definition}[Black Box Inspection Problem]
Given a black box predictor $b$ and a dataset $D = \{ X, Y \}$, the \emph{black box inspection problem} consists in finding a function $f: (\mathcal{X} \rightarrow \mathcal{Y}) \times (\mathcal{X}^n \times \mathcal{Y}^n) \rightarrow \mathcal{V}$ which takes as input a black box $b$ and a dataset $D$, and returns a visual representation of the behavior of the black box, $f(b,D) = v$ with $V$ being the set of all possible representations.
\end{definition}

For example, the function $f$ may be a technique based on sensitivity analysis that, by observing the changing occurring in the predictions when varying the input of $b$, returns a set of visualizations (e.g, partial dependence plots \cite{krause2016interacting}, or variable effect characteristic curve \cite{cortez2013using}) highlighting the feature importance for the predictions.  

\begin{figure}[!tb]
\centering
	\includegraphics[trim = 0mm 0mm 0mm 0mm, clip, width=0.48\linewidth]{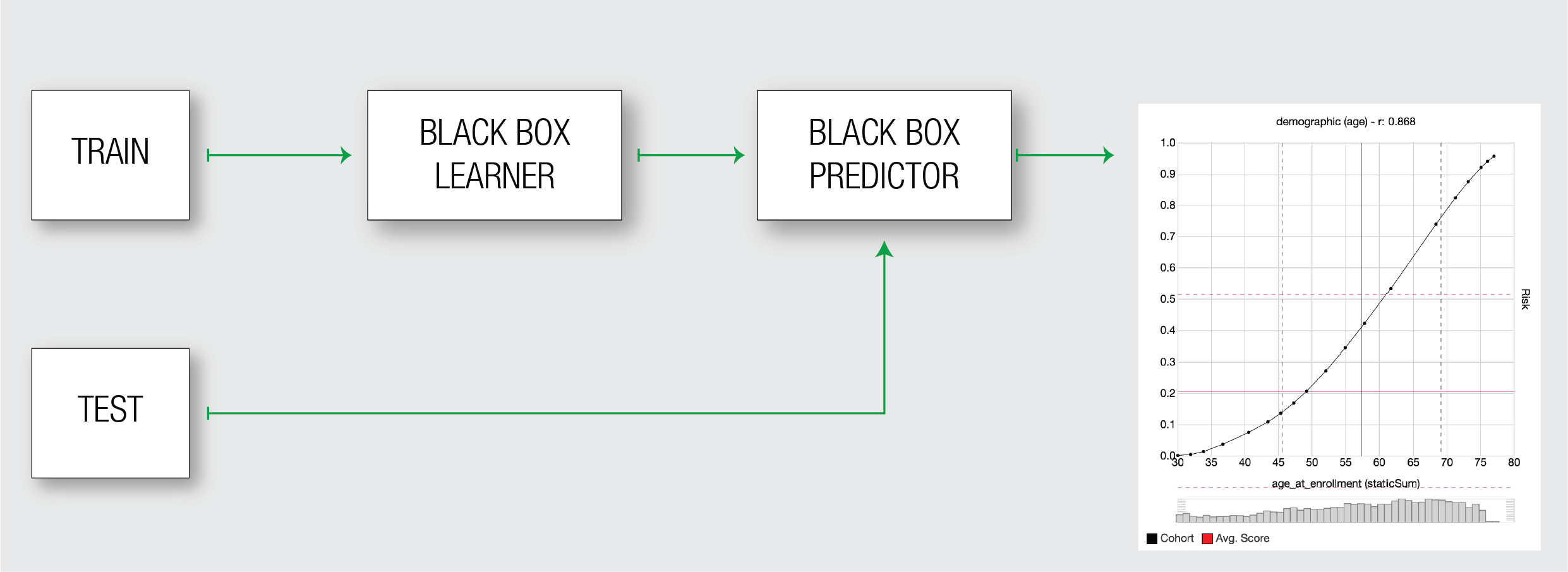}
		\caption{Black Box Inspection Problem.}
	\label{fig:prob_inspection}
\end{figure}

\subsection*{Transparent Box Design Problem} 
\label{sec:transparent_design_problem}
Given a classification problem the \textit{transparent box design problem} consists in providing a model which is locally or globally interpretable on its own. 

\begin{definition}[Transparent Box Design Problem]
Given a dataset $D = \{ X, Y \}$, the \emph{transparent box design problem} consists in finding a learning function $\mathscr{L}_c : (\mathcal{X}^{n \times m} \times \mathcal{Y}) \rightarrow (\mathcal{X}^m \rightarrow \mathcal{Y})$ which takes as input the dataset $D = \{ X, Y \}$ and returns a (locally or globally) comprehensible predictor $c$, i.e., $\mathscr{L}_c(D) = c$. 
This implies that there exists a local explanator function $\varepsilon_l$ or a global explanator function $\varepsilon_g$ (defined as before) that takes as input the comprehensible predictor $c$ and returns a human understandable explanation $e \in \mathcal{E}$ or a set of explanations $E$. 
\end{definition}

For example, the functions $\mathscr{L}_c$ and $c$ may be the decision tree learner and 
predictor respectively, while the global explanator $\varepsilon_g$ may return as explanation a system for following the choices taken along the various branches of the tree, and $\varepsilon_l$ may return a textual representation of the path followed according to the decision suggested by the predictor.

\begin{figure}[!tb]
\centering
	\includegraphics[trim = 0mm 0mm 0mm 0mm, clip, width=0.48\linewidth]{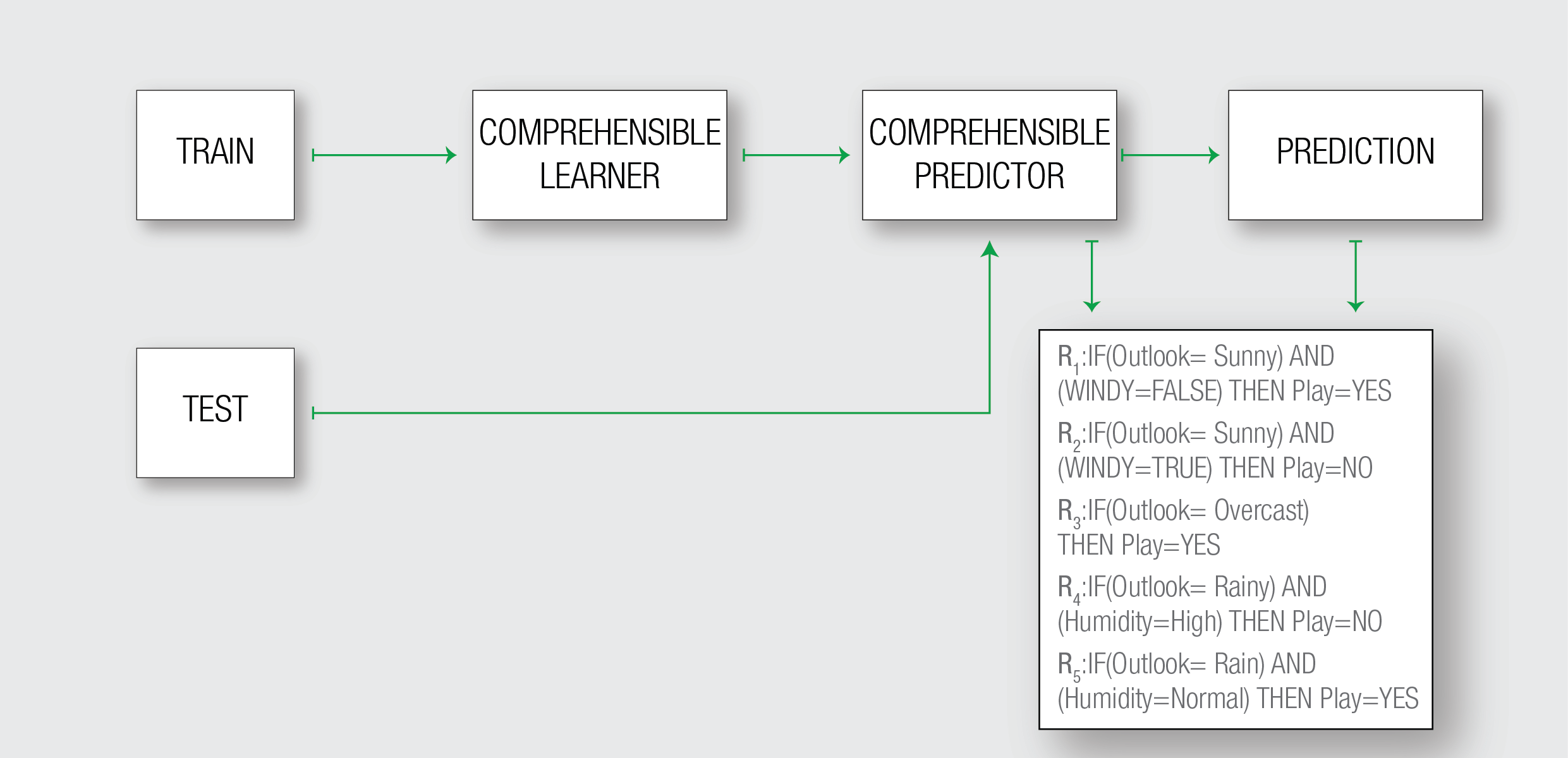}
		\caption{Transparent Box Design Problem.}
	\label{fig:prob_transparent_design}
\end{figure}

\bigskip

Thus, according to our problem definitions, in this survey, when we say that a method is able to \emph{open the black box}, we are referring to one of the following statements: \emph{(i)} it explains the model, \emph{(ii)} it explains the outcome, \emph{(iii)} it can inspect the black box internally, \emph{(iv)} it provides a transparent solution.

\section{Problem And Explanator Based Classification}
\label{sec:classification}
In this survey, we propose a classification based on the type of problem faced and on the explanator adopted to open the black box.
In particular, in our classification we take into account the following features:
\begin{itemize}
    \item the type of \emph{problem} faced (according to the definitions in Section~\ref{sec:problem});
    \item the type of \emph{explanator} adopted to open the black box;
    \item the type of \emph{black box model} that the explanator is able to open;
    \item the type of \emph{data} used as input by the black box model.
\end{itemize}
In each section we group together all the papers that share the same problem definition, while the subsections correspond to the different solutions adopted to develop the explanators.
In turn, in each subsection, we group the papers that try to explain the same type of black box.
Finally, we keep the type of data used by the black box as a feature which is specified for each work analyzed.

We organize the sections discussing the different problems as follows.
In Section~\ref{sec:model_explanation} we analyze the papers presenting approaches to solve the \emph{black box model explanation problem}.
These approaches provide a globally interpretable predictor which is able to mimic the black box.
On the other hand, in Section~\ref{sec:outcome_explanation} are reviewed the methods solving the \emph{black box outcome explanation problem}: the predictor returned is locally interpretable and provides an explanation only for a given record.
In Section~\ref{sec:inspection} we discuss the papers proposing methodologies for \emph{inspecting black boxes}, i.e., not providing a comprehensible predictor but a visualization tool for studying how the black box work internally, and what can happen when a certain input is provided.
Finally, in Section~\ref{sec:transparent_design} we report the papers designing a \emph{transparent} predictor to overcome the ``obscure'' limitation of black boxes.
These approaches try to provide a global or local interpretable model without sacrificing the accuracy of a black box learned to solve the same task.

For each of the sections above, we propose a further categorization with respect to the type of explanator adopted.
This categorization reflects on the papers grouped into the various subsections:
\begin{itemize}
    \item \emph{Decision Tree (DT) or Single Tree}. It is commonly recognized that decision tree is one of the more interpretable and easily understandable models, primarily for global, but also for local, explanations. Indeed, a very widespread technique for opening the black box is the so-called ``single tree approximation''.
    \item \emph{Decion Rules (DR) or Rule Based Explanator}. Decision rules are among the more human understandable techniques. There exist various types of rules (illustrated in Section \ref{sec:recognized_interpretable_models}). They are used to explain the model, the outcome and also for the transparent design. We remark the existence of techniques for transforming a tree into a set of rules.
    \item \emph{Features Importance (FI)}. A very simple but effective solution acting as either global or local explanation consists in returning as explanation the set of features used by the black box together with their weight.
    \item \emph{Salient Mask (SM)}. An efficient way of pointing out what causes a certain outcome, especially when images or texts are treated, consists in using ``masks'' visually highlighting the determining aspects of the record analyzed. They are generally used to explain deep neural networks.
    \item \emph{Sensitivity Analysis (SA)}. It consists of evaluating the uncertainty in the outcome of a black box with respect to different sources of uncertainty in its inputs.
    It is generally used to develop visual tools for black box inspection.
    \item \emph{Partial Dependence Plot (PDP)}. These plots help in visualizing and understanding the relationship between the outcome of a black box and the input in a reduced feature space.
    \item \emph{Prototype Selection (PS)}. This explanator consists in returning, together with the outcome, an example very similar to the classified record, in order to make clear which criteria the prediction was returned. A prototype is an object that is representative of a set of similar instances and is part of the observed points, or it is an artifact summarizing a subset of them with similar characteristics.
    \item \emph{Neurons Activation (NA)}. The inspection of neural networks and deep neural network can be carried out also by observing which are the fundamental neurons activated with respect to particular input records.
\end{itemize}

In the following, we list all the black boxes opened in the reviewed papers.
These black boxes are all supervised learning algorithm designed to solve a classification problem \cite{tan2006introduction}.
\begin{itemize}
    \item \emph{Neural Network (NN)}. Inspired by biological neural networks, artificial neural networks learn to do tasks by considering examples. A NN is formed by a set of connected neurons. Each link between neurons can transmit a signal. The receiving neuron can process the signal and then transmit to downstream neurons connected to it.
    Typically, neurons are organized in layers. Different layers perform different transformations on their inputs. Signals travel from the input layer, to the output layer, passing through the hidden layer(s) in the middle multiple times. Neurons and connections may also have a weight that varies as learning proceeds, which can increase or decrease the strength of the signal that it sends.  
    \item \emph{Tree Ensemble (TE)}. Ensemble methods combine more than one learning algorithm to improve the predictive power of any of the single learning algorithms that they combines. Random forests, boosted trees and tree bagging are examples of TEs. They combine the predictions of different decision trees each one trained on an independent subset of the input data.
    \item \emph{Support Vector Machine (SVM)}. Support Vector Machines utilize a subset of the training data, called support vectors, to represent the decision boundary. A SVN is a classifier that searches for hyperplanes with the largest margin for the decision boundary.
    \item \emph{Deep Neural Network (DNN)}. A DNN is a NN that can model complex non-linear relationship with multiple hidden layers. A DNN architecture is formed by a composition of models expressed as a layered combination of basic units. In DNNs the data typically flows from the input to the output layer without looping back. The most used DNN are Recurrent Neural Networks (RNNs). A peculiar component of RNNs are Long Short-Term Memory (LSTM) nodes which are particularly effective for language modeling. On the other hand, in image processing Convolutional Neural Networks (CNNs) are typically used.
\end{itemize}

Moreover, recently \emph{agnostic} approaches for explaining black boxes are being developed.
An \emph{Agnostic Explanator (AGN)} is a comprehensible predictor which is not tied to a particular type of black box, explanation or data type. 
In other words, in theory, an agnostic predictor can explain indifferently a neural network or a tree ensemble using a single tree or a set of rules.
Since only a few approaches in the literature describe themselves to be \emph{fully} agnostic, and since the principal task is to explain a black box predictor, in this paper, if not differently specified, we term \emph{agnostic} the approaches defined to explain any type of black box.

The types of data used as input of black boxes analyzed in this survey are the following:
\begin{itemize}
    \item \emph{Tabular (TAB)}. With tabular data, we indicate any classical dataset in which every record shares the same set of features and each feature is either numerical, categorical or boolean.
    \item \emph{Image (IMG)}. Many black boxes work with labeled images. These images can be treated as they are by the black box or can be preprocessed (e.g, re-sized in order to have all the same dimensions).
    \item \emph{Text (TXT)}. As language modeling is one of the tasks most widely assessed nowadays together with image recognition, labeled datasets of text are generally used for tasks like spam detection or topic classification.
\end{itemize}
In data mining and machine learning many other types of data are also used like sequences, networks, mobility trajectories, etc.
However, they are not used as input in the methods of the papers proposing a solution for opening the black box.

Table~\ref{tab:summary} lists the methods for opening and explaining black boxes and summarizes the various fundamental features and characteristics listed so far, together with additional information that we believe could be useful for the reader. 
The columns \emph{Examples}, \emph{Code} and \emph{Dataset} indicates if any kind of example of explanation is shown in the paper, and if the source code and the dataset used in the experiments are publicly available, respectively.
The columns \emph{General} and \emph{Random} are discussed in the following section.
We point out that Table~\ref{tab:summary} reports the main references only, while existing extensions or derived works are discussed in the survey.
Table~\ref{tab:summary_legend} reports the legend of Table~\ref{tab:summary}, i.e., the expanded acronym and the meaning of the features in Table~\ref{tab:summary}.
Moreover, in order to provide the reader with a useful tool to find a particular set of papers with determined characteristics, Appendix \ref{sec:supplementary_material} provides Tables \ref{tab:problem}, \ref{tab:explanator} and \ref{tab:blackbox}, in which are reported the list of the papers with respect to each problem, explanator and black box, respectively.


    
\begin{table}
	\begin{adjustwidth}{-3.5cm}{}
    \centering
    \setlength{\tabcolsep}{1mm}
    \caption{Summary of methods for opening and explaining black boxes. 
    }
  \begin{tabular}{ccccccccccccc}
    \hline
    \rotatebox[origin=c]{55}{\textbf{Name}} & \rotatebox[origin=c]{55}{\textbf{Ref.}}  & \rotatebox[origin=c]{55}{\textbf{Authors}} & \rotatebox[origin=c]{55}{\textbf{Year}}  & \rotatebox[origin=c]{55}{\textbf{Problem}} & \rotatebox[origin=c]{55}{\textbf{Explanator}} & \rotatebox[origin=c]{55}{\textbf{Black Box}} & \rotatebox[origin=c]{55}{\textbf{Data Type}}  & \rotatebox[origin=c]{55}{\textbf{General}} & \rotatebox[origin=c]{55}{\textbf{Random}} & \rotatebox[origin=c]{55}{\textbf{Examples}} & \rotatebox[origin=c]{55}{\textbf{Code}}  & \rotatebox[origin=c]{55}{\textbf{Dataset}} \\
    
    \hline
    \rowcolor{gray!15} Trepan & \cite{craven1996extracting} & Craven et al. & 1996  & Model Expl. & DT & NN & TAB & \checkmark & & & & \checkmark \\
 -  & \cite{krishnan1999extracting} & Krishnan et al. & 1999  & Model Expl. & DT & NN & TAB & \checkmark & & \checkmark & & \checkmark \\
\rowcolor{gray!15}  DecText & \cite{boz2002extracting} & Boz & 2002  & Model Expl. & DT & NN & TAB & \checkmark & \checkmark & & & \checkmark \\
 GPDT  & \cite{johansson2009evolving} & Johansson et al. & 2009  & Model Expl. & DT & NN & TAB & \checkmark & \checkmark & \checkmark & & \checkmark \\
\rowcolor{gray!15}  Tree Metrics & \cite{chipman1998making} & Chipman et al. & 1998  & Model Expl. & DT & TE & TAB & & & & & \checkmark \\
 CCM & \cite{domingos1998knowledge} & Domingos et al. & 1998  & Model Expl. & DT & TE & TAB & \checkmark & \checkmark & & & \checkmark \\
\rowcolor{gray!15}  -  & \cite{gibbons2013cad} & Gibbons et al. & 2013  & Model Expl. & DT & TE & TAB & \checkmark & \checkmark & & &  \\
 STA & \cite{zhou2016interpreting} & Zhou et al. & 2016  & Model Expl. & DT & TE & TAB & & \checkmark & & &  \\
\rowcolor{gray!15}  CDT & \cite{schetinin2007confident} & Schetinin et al. & 2007  & Model Expl. & DT & TE & TAB & & & \checkmark & &  \\
 -  & \cite{hara2016making} & Hara et al. & 2016  & Model Expl. & DT & TE & TAB & & \checkmark & \checkmark & & \checkmark \\
\rowcolor{gray!15}  TSP & \cite{tan2016tree} & Tan et al. & 2016  & Model Expl. & DT & TE & TAB & & & & & \checkmark \\
 Conj Rules & \cite{craven1994using} & Craven et al. & 1994  & Model Expl. & DR & NN & TAB & & \checkmark & & &  \\
\rowcolor{gray!15}  G-REX & \cite{johansson2003rule} & Johansson et al. & 2003  & Model Expl. & DR & NN & TAB & \checkmark & \checkmark & \checkmark & &  \\
 REFNE & \cite{zhou2003extracting} & Zhou et al. & 2003  & Model Expl. & DR & NN & TAB & \checkmark & \checkmark & \checkmark & & \checkmark \\
\rowcolor{gray!15}  RxREN & \cite{augasta2012reverse} & Augasta et al. & 2012  & Model Expl. & DR & NN & TAB & & \checkmark & \checkmark & & \checkmark \\
 SVM+P & \cite{nunez2002rule} & Nunez et al. & 2002  & Model Expl. & DR & SVM & TAB & & & \checkmark & & \checkmark \\
\rowcolor{gray!15}  -  & \cite{fung2005rule} &  Fung et al. & 2005  & Model Expl. & DR & SVM & TAB & & & \checkmark & & \checkmark \\
 inTrees & \cite{deng2014interpreting} & Deng  & 2014  & Model Expl. & DR & TE & TAB & & & \checkmark & & \checkmark \\
\rowcolor{gray!15}  -  & \cite{lou2013accurate} & Lou et al. & 2013  & Model Expl. & FI & AGN & TAB & \checkmark & & \checkmark & \checkmark & \checkmark \\
 GoldenEye & \cite{henelius2014peek} & Henelius et al. & 2014  & Model Expl. & FI & AGN & TAB & \checkmark & \checkmark & \checkmark & \checkmark & \checkmark \\
\rowcolor{gray!15}  PALM  & \cite{krishnan2017palm} & Krishnan et al. & 2017  & Model Expl. & DT & AGN & ANY & \checkmark & & \checkmark & & \checkmark \\
 -  & \cite{tolomei2017interpretable} & Tolomei et al. & 2017  & Model Expl. & FI & TE & TAB & & & \checkmark & &  \\
\rowcolor{gray!15}  -  & \cite{xu2015show} & Xu et al. & 2015  & Outcome Expl. & SM & DNN & IMG & & & \checkmark & \checkmark & \checkmark \\
 -  & \cite{fong2017interpretable} & Fong et al. & 2017  & Outcome Expl. & SM & DNN & IMG & & & \checkmark & & \checkmark \\
\rowcolor{gray!15}  CAM & \cite{zhou2016learning} & Zhou et al. & 2016  & Outcome Expl. & SM & DNN & IMG & & & \checkmark & \checkmark & \checkmark \\
 Grad-CAM & \cite{selvaraju2016grad} & Selvaraju et al. & 2016  & Outcome Expl. & SM & DNN & IMG & & & \checkmark & \checkmark & \checkmark \\
\rowcolor{gray!15}  -  & \cite{lei2016rationalizing} & Lei et al. & 2016  & Outcome Expl. & SM & DNN & TXT & & & \checkmark & & \checkmark \\
 LIME  & \cite{ribeiro2016nothing} & Ribeiro et al. & 2016  & Outcome Expl. & FI & AGN & ANY & \checkmark & \checkmark & \checkmark & \checkmark & \checkmark \\
\rowcolor{gray!15}  MES & \cite{turner2016model} & Turner et al. & 2016  & Outcome Expl. & DR & AGN & ANY & \checkmark & & \checkmark & & \checkmark \\
 NID & \cite{olden2002illuminating} & Olden et al. & 2002  & Inspection & SA & NN & TAB & & & \checkmark & &  \\
\rowcolor{gray!15}  GDP & \cite{baehrens2010explain} & Baehrens & 2010  & Inspection & SA & AGN & TAB & \checkmark & & \checkmark & & \checkmark \\
 IG & \cite{sundararajan2017axiomatic} & Sundararajan & 2017  & Inspection & SA & DNN & ANY & & & \checkmark & & \checkmark \\
\rowcolor{gray!15}  VEC & \cite{cortez2011opening} & Cortez et al. & 2011  & Inspection & SA & AGN & TAB & \checkmark & & \checkmark & & \checkmark \\
 VIN & \cite{hooker2004discovering} & Hooker & 2004  & Inspection & PDP & AGN & TAB & \checkmark & & \checkmark & & \checkmark \\
\rowcolor{gray!15}  ICE & \cite{goldstein2015peeking} & Goldstein et al. & 2015  & Inspection & PDP & AGN & TAB & \checkmark & & \checkmark & \checkmark & \checkmark \\
 Prospector & \cite{krause2016interacting} &  Krause et al. & 2016  & Inspection & PDP & AGN & TAB & \checkmark & & \checkmark & & \checkmark \\
\rowcolor{gray!15}  Auditing & \cite{adler2016auditing} & Adler et al. & 2016  & Inspection & PDP & AGN & TAB & \checkmark & & \checkmark & \checkmark & \checkmark \\
 OPIA  & \cite{adebayo2016iterative} & Adebayo et al. & 2016  & Inspection & PDP & AGN & TAB & \checkmark & & \checkmark & &  \\
\rowcolor{gray!15}  -  & \cite{yosinski2015understanding} & Yosinski et al. & 2015  & Inspection & NA & DNN & IMG & & & \checkmark & & \checkmark \\
 TreeView & \cite{thiagarajan2016treeview} & Thiagarajan et al. & 2016  & Inspection & DT & DNN & TAB & & & \checkmark & & \checkmark \\
\rowcolor{gray!15}  IP & \cite{shwartz2017opening} & Shwartz et al. & 2017  & Inspection & NA & DNN & TAB & & & \checkmark & &  \\
 -  & \cite{radford2017learning} & Radford & 2017  & Inspection & NA & DNN & TXT & & & \checkmark & &  \\
\rowcolor{gray!15}  CPAR  & \cite{yin2003cpar} & Yin et al. & 2003  & Transp. Design & DR & -  & TAB & & & & & \checkmark \\
 FRL & \cite{wang2015falling} & Wang et al. & 2015  & Transp. Design & DR & -  & TAB & & & \checkmark & \checkmark & \checkmark \\
\rowcolor{gray!15}  BRL & \cite{letham2015interpretable} & Letham et al. & 2015  & Transp. Design & DR & -  & TAB & & & \checkmark & &  \\
 TLBR  & \cite{su2015interpretable} & Su et al. & 2015  & Transp. Design & DR & -  & TAB & & & \checkmark & & \checkmark \\
\rowcolor{gray!15}  IDS & \cite{lakkaraju2016interpretable} & Lakkaraju et al. & 2016  & Transp. Design & DR & -  & TAB & & & \checkmark & &  \\
 Rule Set & \cite{wang2016bayesian} & Wang et al. & 2016  & Transp. Design & DR & -  & TAB & & & \checkmark & \checkmark & \checkmark \\
\rowcolor{gray!15}  1Rule & \cite{malioutov2017learning} & Malioutov et al. & 2017  & Transp. Design & DR & -  & TAB & & & \checkmark & & \checkmark \\
 PS & \cite{bien2011prototype} & Bien et al. & 2011  & Transp. Design & PS & -  & ANY & & & \checkmark & & \checkmark \\
\rowcolor{gray!15}  BCM & \cite{kim2014bayesian} & Kim et al. & 2014  & Transp. Design & PS & -  & ANY & & & \checkmark & & \checkmark \\
 -  & \cite{mahendran2015understanding} & Mahendran et al. & 2015  & Transp. Design & PS & -  & IMG & & & \checkmark & \checkmark & \checkmark \\
\rowcolor{gray!15}  -  & \cite{kononenko2010efficient} & Kononenko et al. & 2010  & Transp. Design & FI & -  & TAB & & & \checkmark & & \checkmark \\
 OT-SpAMs & \cite{wang2015trading} & Wang et al. & 2015  & Transp. Design & DT & -  & TAB & & & \checkmark & \checkmark & \checkmark \\
    \hline
    \end{tabular}%
    \label{tab:summary}
    \end{adjustwidth}
    \end{table}

  
\begin{table}
    \centering
    \setlength{\tabcolsep}{1mm}
\caption{Legend of Table ~\ref{tab:summary}.  In the following are described the features reported and the abbreviations adopted.}
    \begin{tabular}{cp{11cm}}
    \hline
    \textbf{Feature} & \textbf{Description} \\
    \hline
    \emph{Problem} & Model Explanation, Outcome Explanation, Black Box Inspection, Transparent Design \\
    \emph{Explanator} & DT - Decision Tree, DR - Decision Rules, FI - Features Importance, {SM - Saliency Masks,} SA - Sensitivity Analysis, PDP - Partial Dependence Plot, NA - Neurons Activation, \mbox{PS - Prototype Selection} \\
    \emph{Black Box} & NN - Neural Network, TE - Tree Ensemble, SVM - Support Vector Machines, \mbox{DNN - Deep Neural Network}, AGN - AGNostic black box \\
    \emph{Data Type} & TAB - TABular, IMG - IMaGe, TXT - TeXT, ANY - ANY type of data \\
    \emph{General} & Indicates if an explanatory approach can be generalized for every black box, i.e., it does not consider peculiarities of the black box to produce the explanation \\
    \emph{Random} & Indicates if any kind of random perturbation or permutation of the original dataset is required for the explanation \\
    \emph{Examples} & Indicates if example of explanations are shown in the paper \\
    \emph{Code} & Indicates if the source code is available \\
    \emph{Dataset} & Indicates if the datasets used in the experiments are available \\
    \hline
    \end{tabular}
    \label{tab:summary_legend}
    %
\end{table}

\subsection*{Reverse Engineering: A Common Approach For Understanding The Black Box}
\begin{figure}[!tb]
\centering
	\includegraphics[trim = 0mm 0mm 0mm 0mm, clip, width=0.48\linewidth]{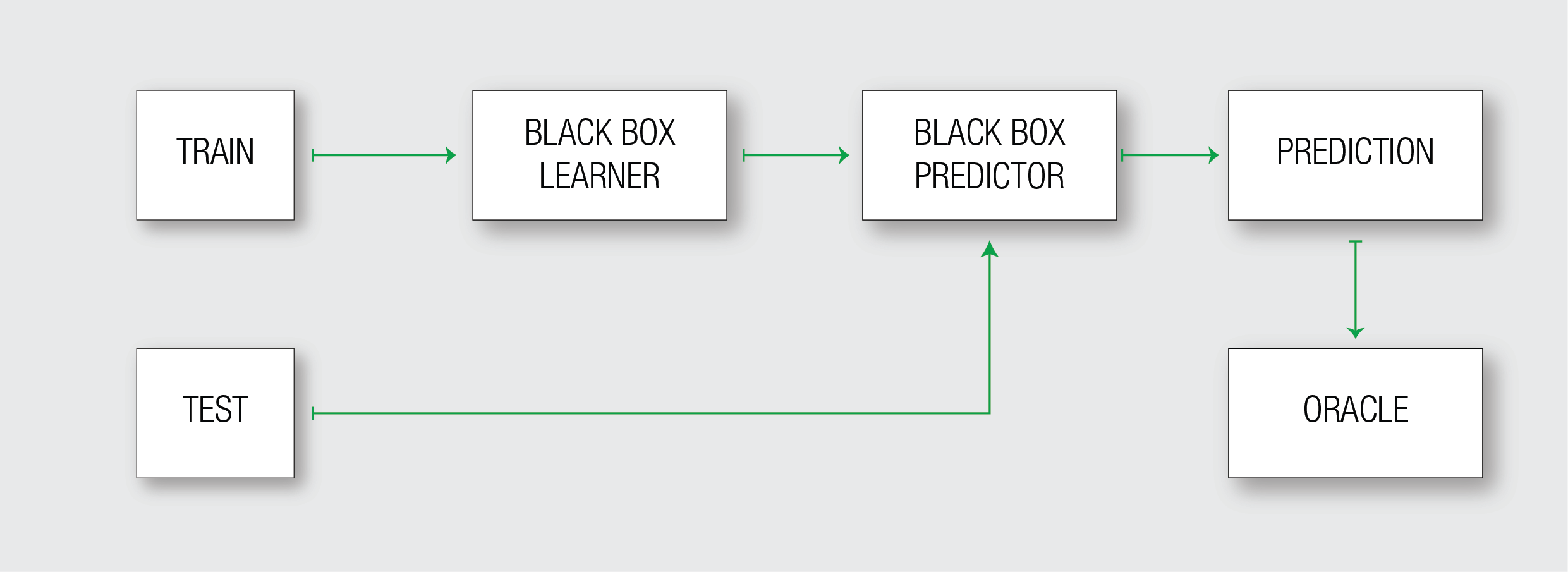}
		\caption{Reverse Engineering approach: the learned black box predictor is queried with a test dataset to produce an oracle which associate to each record a label which is not real but assigned by the black box.}
	\label{fig:reverse_eng}
\end{figure}

Before proceeding in the detailed analysis and classification of papers proposing method $f$ for understanding black boxes $b$, we present in this section the most largely used approach to solve the black box model and outcome explanation problems and the black box inspection problem.
We refer to this approach as \emph{reverse engineering} because the black box predictor $b$ is queried with a certain test dataset in order to create an \emph{oracle} dataset that in turn will be used to train the comprehensible predictor (see Figure~\ref{fig:reverse_eng}).
The name reverse engineering comes from the fact that we can only observe the input and output of the black box.

With respect to the black box model and outcome explanation problems, the possibility of action tied with this approach relies on the choice of adopting a particular type of comprehensible predictor, and in the possibility of querying the black box with input records created in a controlled way and/or by using \emph{random perturbations} of the initial train or test dataset.
Regarding the random perturbations of the input used to feed the black box, it is important to recall that recent studies discovered that DNN built for classification problems on texts and images can be easily fooled (see Section \ref{sec:motivations}).
Not human perceptible changes in an image can lead a DNN to label the record as something else.
Thus, according to these discoveries, the methods treating images or text, in theory, should not be enabled to use completely random perturbations of their input.
However, this is not always the case in practice \cite{ribeiro2016should}.

\begin{figure}[!tb]
\centering
	\includegraphics[trim = 0mm 0mm 0mm 0mm, clip, width=0.47\linewidth]{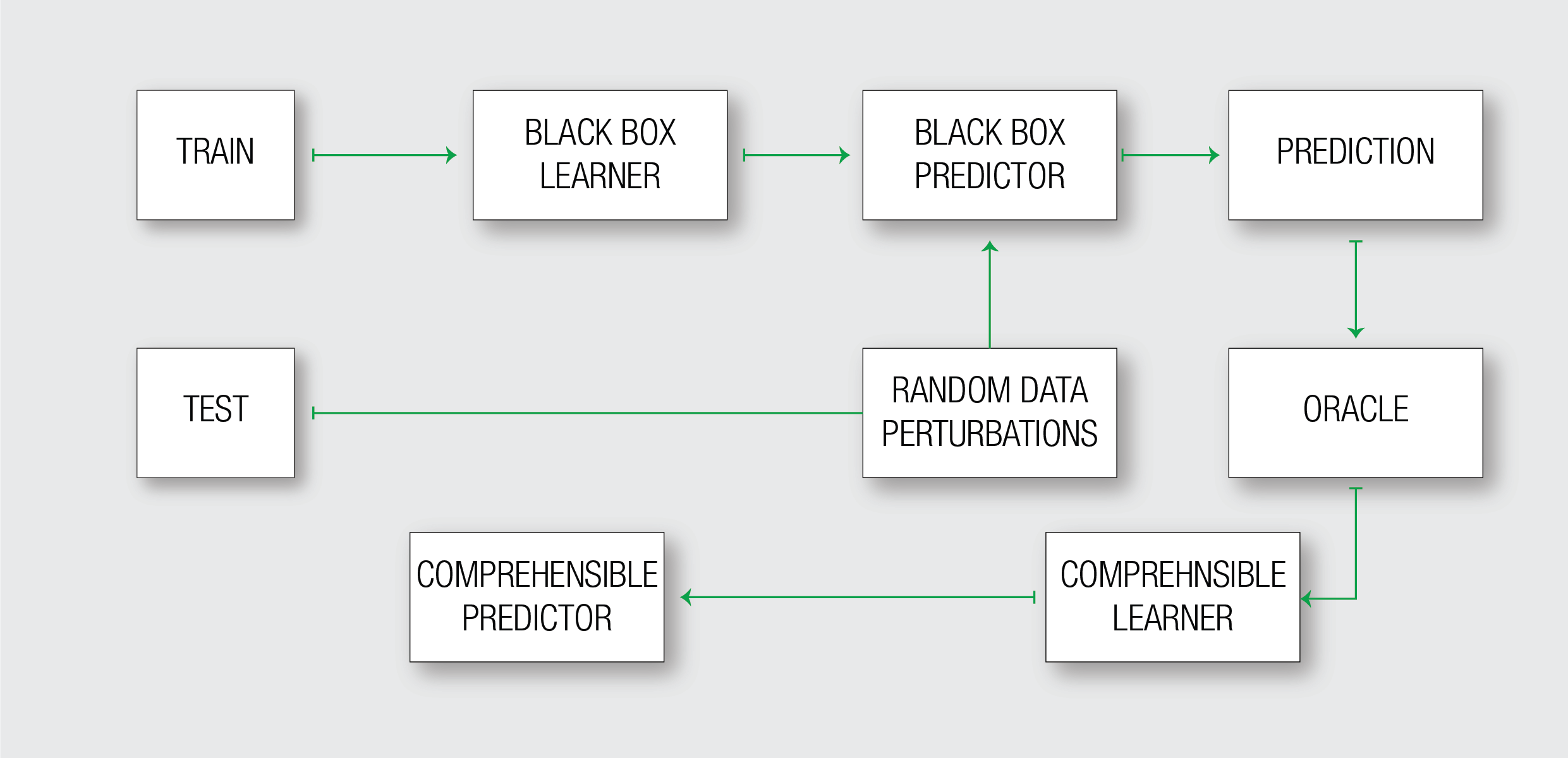}
	\hspace{5mm}
	\includegraphics[trim = 0mm 0mm 0mm 0mm, clip, width=0.47\linewidth]{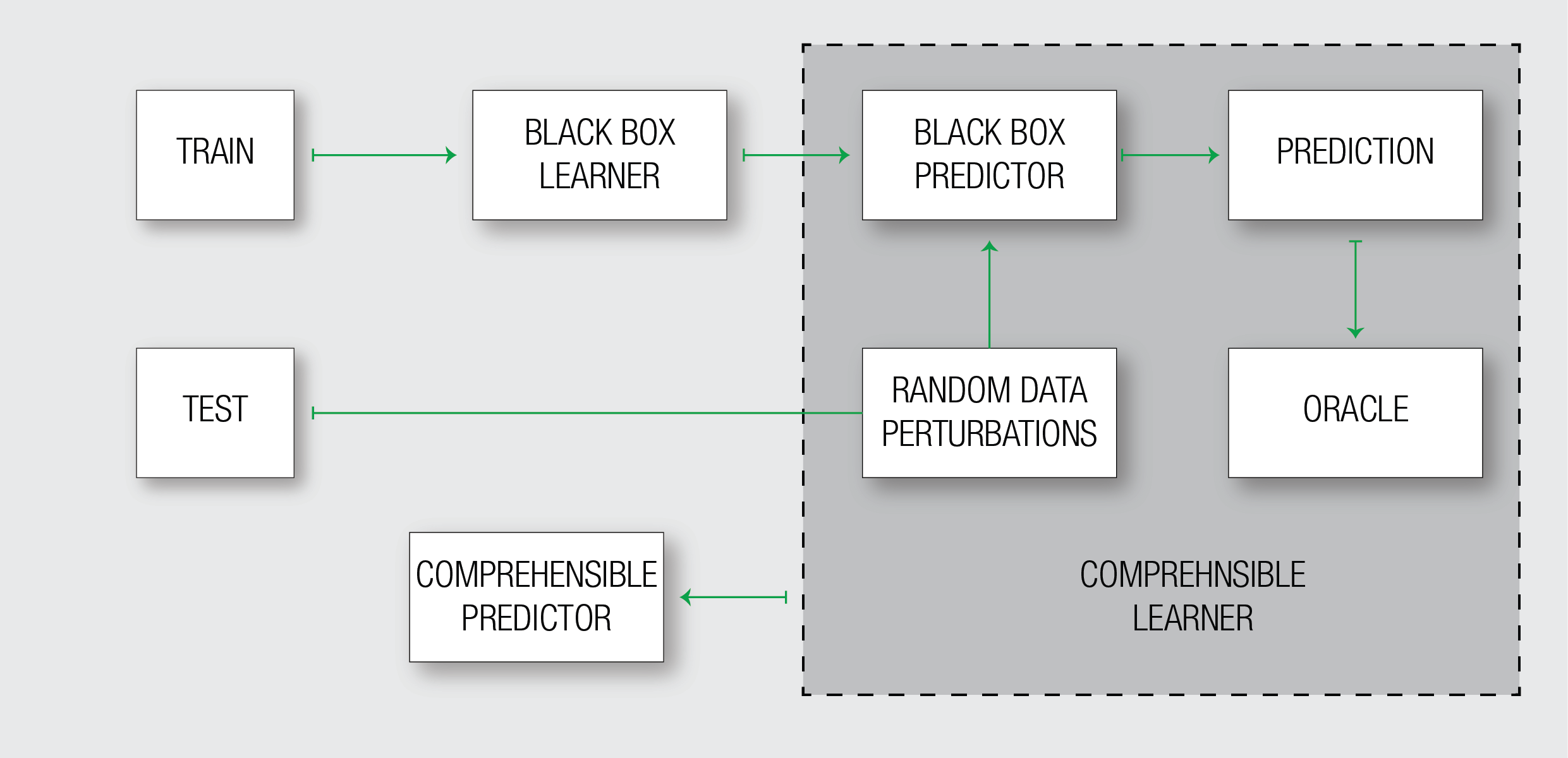}
		\caption{\emph{(Left)} Generalizable reverse engineering approach: internal peculiarities of the black box are not exploited to build the comprehensible predictor. \emph{(Right)} Not Generalizable reverse engineering approach: the comprehensible predictor is the result of a procedure involving internal characteristics of the black box.}
	\label{fig:gen_nogen}
\end{figure}

Such reverse engineering approach can be classified as \emph{generalizable} or not (or pedagocial vs. decompsitional as described in \cite{martens2007comprehensible}).
We say that an approach is generalizable when a purely reverse engineering procedure is followed, i.e., the black box is only queried with different input records to obtain an oracle used for learning the comprehensible predictor (see Figure~\ref{fig:gen_nogen}-\emph{(left)}).
In other words, internal peculiarities of the black box are not exploited to build the comprehensible predictor.
Thus, if an approach is generalizable, even though it is presented to explain a particular type of black box, in reality, it can be used to interpret any kind of black box predictor.
That is, it is an agnostic approach for interpreting black boxes.
On the other hand, we say that an approach is not generalizable if it can be used to open only that particular type of black box for which it was designed for (see Figure~\ref{fig:gen_nogen}-\emph{(right)}).
For example, if an approach is designed to interpret random forest and internally use a concept of distance between trees, then such approach can not be utilized to explain predictions of a NN.
A not generalizable approach can not be black box agnostic.

In Table~\ref{tab:summary} we keep track of these aspects with the two features \emph{General} and \emph{Random}.
With \emph{General} we indicate if an explanatory approach can be generalized for every black box, while with \emph{Random} we indicate if any kind of random perturbation or permutation of the original dataset is used by the explanatory approach.
    
\bigskip
In light of these concepts, as the reader will discover below, a further classification not explicitly indicated emerges from the analysis of these papers.
This fact can be at the same time a strong point or a weakness of the current state of the art.
Indeed, we highlight that the works for opening the black box are realized for two cases.
The first (larger) group contains approaches proposed to tackle a particular problem (e.g., medical cases) or to explain a particular type of black box, that is, the solutions are specific for the problem instance.
The second group contains general purpose solutions that try to be general as much as possible and propose agnostic and generalizable solutions.

\section{Solving the Black Box Model Explanation Problem}
\label{sec:model_explanation}
In this section we review the methods for opening the black box facing the \emph{black box model explanation problem} (see Section~\ref{sec:model_explanation_problem}).
That is, the proposed methods provide globally interpretable models which are able to mimic the behavior of black boxes and which are also understandable by humans. 
We recognized different groups of approaches.
In Section~\ref{sec:model_explanation_tree} we analyze the proposals using a decision tree as explanator, while in Section~\ref{sec:model_explanation_rule} they use rules.
Section~\ref{sec:model_explanation_agnostic} describes the methods which are designed to work with any type of black box.
Finally, Section~\ref{sec:model_explanation_others} contains the remaining ones.

\subsection{Explanation via Single Tree Approximation}
\label{sec:model_explanation_tree}
In this section we present a set of works addressing the \emph{black box model explanation problem} by implementing in different ways the function $f$.
All the following works adopt a decision tree as comprehensible global predictor $c_g$, and consequently represent the explanation $\varepsilon_g$ with the decision tree itself.
Moreover, we point out that all the methods presented in this section work on tabular data.

\subsubsection{Explanation of Neural Networks.}
The following papers describe the implementation of functions $f$ which are able to interpret a black box $b$ consisting in a \emph{Neural Network (NN)} \cite{tan2006introduction} with a comprehensible global predictor $c_g$ consisting in a decision tree.
In these works, the NNs are considered black-boxes, i.e., the only interface permitted is presenting an input $x$ to the neural network $b$ and obtaining the outcome $\hat{y}$. The final goal is to comprehend how the neural networks behave by submitting to it a large set of instances and analyzing their different predictions. 

\smallskip

Single tree approximations for NNs were first presented in 1996 by Craven et al. \cite{craven1996extracting}. 
The comprehensible representations of the neural network $b$ is returned by \emph{Trepan} which is the implementation of function $f$.
Trepan queries the neural network $b$ to induce a decision tree $c_g$ approximating the concepts represented by the networks by maximizing the gain ratio \cite{tan2006introduction} together with an estimation of the current model fidelity.
Another advantage of Trepan with respect to common tree classifiers like ID3 or C4.5 \cite{tan2006introduction} is that, thanks to the black box $b$, it can use as many instances as desired for each split, so that also the node splits near to the bottom of the tree are realized using a considerable amount of training data.

In \cite{krishnan1999extracting}, Krishnan et al. present a three step method $f$. 
The first step generates a sort of ``prototype'' for each target class in $Y$ by using genetic programming to query the trained neural network $b$.
The input features dataset $X$ is exploited for constraining the prototypes.
The second step selects the best prototypes for inducing the learning of the decision tree $c_g$ in the third step. 
This approach leads to get more understandable and smaller decision trees starting from smaller data sets.

In \cite{boz2002extracting}, Boz describes \emph{DecText}, another procedure that uses a decision tree $c$ to explain neural network black boxes $b$.
The overall procedure recalls Trepan \cite{craven1996extracting} with the innovation of four splitting methods aimed at finding the most relevant features during the tree construction.
Moreover, since one of the main purposes of the tree is to maximize the fidelity while keeping the model simple, a fidelity pruning strategy to reduce the tree size is defined.
A set of random instances are generated. 
Then, starting from the bottom of the tree, for each internal node a leaf is created with the majority label using the labeling of the random instances.
If the fidelity of the new tree overtakes the old one, than the maximum fidelity and the tree are updated.

In \cite{johansson2009evolving} Johansson et al. use \emph{Genetic Programming} to evolve Decision Trees (the comprehensible global predictor $c_g$), in order to mimic the behavior of a neural network ensemble (the black box $b$).
The dataset $D$ used by genetic programming (implementing function $f$) consists of a lot of different combinations of the original data and oracle data labeled by $b$. 
The paper shows that trees based only on original training data have the worst performance in terms of accuracy in the test data, while the trees evolved using both the oracle guide and the original data produce significantly more accurate trees $c_g$.

\medskip
We underline that, even though these approaches are developed to explain neural networks, since peculiarities of the neural networks are not used by $f$, which uses $b$ only as an oracle, these approaches can be potentially adopted as agnostic explanators, i.e., they can be used to open any kind of black box and represent it with a single tree.

\subsubsection{Explanation of Tree Ensembles.}
Richer collections of trees provide higher performance and less uncertainty in the prediction.
On the other hand, it is generally difficult to make sense of the resultant forests.
The papers presented in this section describe functions $f$ for approximating a black box model $b$ consisting in \emph{Tree Ensembles (TE)} \cite{tan2006introduction} (e.g. random forests) with a global comprehensible predictor $c_g$ in the form of a decision tree, and explanation $\varepsilon_g$ as a the decision tree as before.

\smallskip

Unlike previous works, the tree ensembles are not only viewed as black boxes, but also some of their internal features are used to derive the global comprehensible model $c_g$. 
For example, Chipman et al., in \cite{chipman1998making} observe that although hundreds of distinct trees are identified by \emph{random forests}, in practice, many of them generally differ only by few nodes. 
In addition, some trees may differ only in the topology, but use the same partitioning of the feature space $\mathcal{X}$.
The paper proposes several measures of dissimilarity for trees.
Such measures are used to summarize forest of trees through clustering, and finally use archetypes of the associated clusters as model explanation.
Here, $f$ corresponds to the clustering procedure, and the global comprehensible predictor $c_g$ is the set of tree archetypes minimizing the distance among all the trees in each cluster.
In this approach, $f$ does not extend the input dataset $D$ with random data.

On the other hand, random data enrichment and model combination are the basis for the \textit{Combined Multiple Model (CCM)} procedure $f$ presented in \cite{domingos1998knowledge}. 
Given the tree ensemble black box $b$, it first modifies $n$ times the input dataset $D$ and learns a set of $n$ black boxes $b_i$ $\forall i = 1, \dots, n$, and then it randomly generates data record $x$ which are labeled using a \emph{combination} (e.g. bagging) of the $n$ black boxes $b_i$, i.e., $C_{b_1, \dots, b_n }(x) = \hat{y}$. In this way, the training dataset $D = D \cup \{ x, \hat{y} \}$ is increased.
Finally, it builds the global comprehensible model $c_g$ as a decision tree (C4.5 \cite{quinlan1993c4}) on the enriched dataset $D$.
Since it is not exploiting particular features of the tree ensemble $b$, also this approach can be generalized with respect to the black box $b$.
In line with \cite{domingos1998knowledge}, the authors of \cite{gibbons2013cad} generate a very large artificial dataset $D$ using the prediction of the random forest $b$, then explain $b$ by training a decision tree $c_g$ on this artificial dataset in order to mime the behavior of the random forest. 
Finally, they improve the comprehensibility of $c_g$ by cutting the decision tree with respect to a human understandable depth (i.e., from 6 to 11 nodes of depth).
\cite{zhou2016interpreting} proposes \emph{Single Tree Approximation (STA)}, an extension of \cite{gibbons2013cad} which empowers the construction of the final decision tree $c_g$ by using test hypothesis to understand which are the best splits observing the Gini indexes on the trees of the random forest $b$.

Schetinin et al. in \cite{schetinin2007confident} present an approach for the probabilistic interpretation of the black box $b$ \emph{Bayesian decision trees ensembles} \cite{breiman1984classification} through a quantitative evaluation of uncertainty of a \emph{Confident Decision Tree} (CDT) $c_g$.
The methodology $f$ for interpreting $b$ is summarized as follows: \emph{(i)} the classification confidence for each tree in the ensemble is calculated using the training data $D$, \emph{(ii)} the decision tree $c_g$ that covers the maximal number of correct training examples is selected, keeping minimal the amount of misclassifications on the remaining examples by sub-sequentially refining the training dataset $D$.
Similarly to \cite{chipman1998making}, also this explanation method $f$ does not extend the input dataset $D$ with random data and cannot be generalized to other black boxes but can be used only with Bayesian decision tree ensembles.

In \cite{hara2016making}, Hara et al. reinterpret \emph{Additive Tree Models (ATM)} (the black box $b$) using a probabilistic generative model interpretable by humans. 
An interpretable ATM has a sufficiently small number of regions.
Therefore, their aim is to reduce the number of regions in an ATM while minimizing the model error.
To satisfy these requirements, they propose a post processing method $f$ that works as follows.
First, it learns an ATM $b$ generating a number of regions. 
Then, it mimics $b$ using a simpler model (the comprehensible global predictor $c_g$) where the number of regions is fixed as small, e.g., ten. 
In particular, to obtain the simpler model an Expectation Maximization algorithm is adopted \cite{tan2006introduction} minimizing the Kullback-Leibler divergence from the ensemble $b$.

The authors of \cite{tan2016tree} propose \emph{Tree Space Prototype (TSP)}, an approach $f$ for interpretating tree ensembles (the black box $b$) by finding tree prototypes (the comprehensible global predictor $c_g$) in the tree space. 
The main contributions for $f$ are: \emph{(i)} the definition of the \emph{random forest proximity} between trees, and \emph{(ii)} the design of the procedure to extract the tree prototypes used for classification.

\subsection{Explanation via Rule Extraction}
\label{sec:model_explanation_rule}
Another commonly used state of the art interpretable and easily understandable model is the \emph{set of rules}.
When a set of rules describing the logic behind the black box model is returned the interpretability is provided at a global level.
In the following, we present a set of reference works solving the \emph{black box model explanation problem} by implementing in different ways function $f$, and by adopting any kind of \emph{decision rules} as comprehensible global predictor $c_g$.
Hence, the global explanation $\varepsilon_g$ change accordingly to the type of rules extracted by $c_g$.
Similarly to the previous section, also all the methods presented in this section work on tabular data.

\subsubsection{Explanation of Neural Networks.}
The following papers describe the implementation of functions $f$ which are able to interpret a black box $b$ consisting in a \emph{Neural Network (NN)} \cite{tan2006introduction}. 
In the literature already exists a survey specialized on techniques extracting rules from neural networks \cite{andrews1995survey}.
It provides an overview of mechanisms designed to \emph{(i)} insert knowledge into neural networks (knowledge initialization), \emph{(ii)} extract rules from trained NNs (rule extraction), and \emph{(iii)} use NNs to refine existing rules (rule refinement). 
The approaches presented in \cite{andrews1995survey} are strongly dependent on the black box $b$ and on the specific type of decision rules $c_g$.
Thus, they are not generalizable and can not be employed to solve other instances of the black box model explanation problem. 
The survey \cite{andrews1995survey} classifies the methods according to the following criteria:
\begin{itemize}
    \item Expressive power of the extracted rules.
    \item Translucency: that is decompositional, pedagogical and eclectic properties.
    \item Portability of the rule extraction technique.
    \item Quality of the rules extracted. Quality includes accuracy, fidelity, consistency, i.e., different training of the NN extract the rules that lead to the same classification of unseen examples.
    \item Algorithmic complexity.
\end{itemize}

A typical paper analyzed in \cite{andrews1995survey} is \cite{craven1994using} where Craven et al. present a method $f$ to explain the behavior of a neural network $b$ by transforming rule extraction (which is a search problem) into a learning problem.
The original training data $D$ and a randomized extension of it are provided as input to the black box $b$.
If the input $x \in D$ with outcome $\hat{y}$ is not covered by the set of rules, then a \emph{conjunctive} (or m-of-n) rule is formed from $\{x,\hat{y}\}$ considering all the possible antecedents.
The procedure ends when all the target classes have been processed.

In \cite{johansson2004accuracy} Johansson et al. exploit \emph{G-REX} \cite{johansson2003rule}, an algorithm for rule extraction, as function $f$ to explain a neural network $b$.
They use the classical reverse engineering schema where random permutations of the original dataset $D$ are annotated by $b$, and such dataset is used as input by G-REX, which corresponds with $c_g$ in this case.
In particular, G-REX extracts rules by exploiting genetic programming as a key concept.
In subsequent works, the authors show that the proposed methodology $f$ can be also employed to interpret trees ensembles.
\cite{johansson2004truth} extends G-REX for handling regression problems by generating regression trees, and classification problems by generating fuzzy rules.

In \cite{zhou2003extracting} the authors present \emph{REFNE}, an approach $f$ to explain neural network ensembles $b$. 
REFNE uses ensembles for generating instances and then, extracts symbolic rules $c_g$ from those instances. 
REFNE avoids useless discretizations of continuous attributes, by applying a particular discretization leading to discretize different continuous attributes using different intervals.
Moreover, REFNE can also be used as a rule learning approach, i.e., it solves the transparent box design problem (see Section~\ref{sec:transparent_design_problem}).
Also in \cite{augasta2012reverse} Augasta et al. propose 
\emph{RxREN} a rule extraction algorithm $c_g$ which returns the explanation of a trained NN $b$.
The method $f$ works as follows. 
First, it prunes the insignificant input neurons from trained NNs and identifies the  data range necessary to classify the given test instance with a specific class.
Second, using a reverse engineering technique, through RxREN generates the classification rules for each class label exploiting the data ranges previously identified, and improve the initial set of rules by a process that prunes and updates the rules. 

\subsubsection{Explanation of Support Vector Machines.}
The following papers show implementations of functions $f$ for explaining a black box $b$ consisting in a \emph{Support Vector Machine} (SVM) \cite{tan2006introduction} still returning a comprehensible global predictor $c_g$ consisting in a set of decision rules.

The authors of \cite{nunez2002rule} propose the \emph{SVM+Prototypes (SVM+P)} procedure $f$ for rule extraction $c_g$ from support vector machines $b$. 
It works as follows: it first determines the decision function by means of a SVM, then a clustering algorithm is used to find out a prototype vector for each class.
By using geometric methods, these points are joined with the support vectors for defining ellipsoids in the input space that can be transformed into if-then rules. 

Fung et al., in \cite{fung2005rule}, describe as function $f$ an algorithm based on constraint programming for converting linear SVM $b$ (and other hyperplane-based linear classifiers) into a set of non overlapping and interpretable rules $c_g$. These rules are asymptotically equivalent to the original linear SVM.
Each iteration of the algorithm for extracting the rules is designed to solve a constrained optimization problem having a low computational cost.
We underline that this black box explanation solution $f$ is not generalizable and can be employed only for Linear SVM-like black boxes.

In \cite{martens2007comprehensible} the authors propose a qualitative comparison of the explanations returned by techniques for extraction of rules from SVM black boxes (e.g. SVM+P \cite{nunez2002rule}, Fung method \cite{fung2005rule}) against the redefining of methods designed for explaining neural networks, i.e., C4.5 \cite{tan2006introduction}, Trepan \cite{craven1996extracting} and G-REX \cite{johansson2003rule}.
How we anticipated in the previous section, the authors delineate the existence of two type of approaches to extract rules: \emph{pedagogical} and \emph{decompositional}. 
Pedagogical techniques $f$ directly extract rules which relate the inputs and outputs of the predictor (e.g. \cite{craven1996extracting,johansson2003rule}), while decompositional approaches are closely intertwined with the internal structure of the SVM (e.g. \cite{nunez2002rule,fung2005rule}).
We recall that, in Table~\ref{tab:summary} we identify with the term generalizable the pedagogical approaches. 

\begin{figure}[!tb]
\centering
	\includegraphics[trim = 0mm 0mm 0mm 0mm, clip, width=0.4\linewidth]{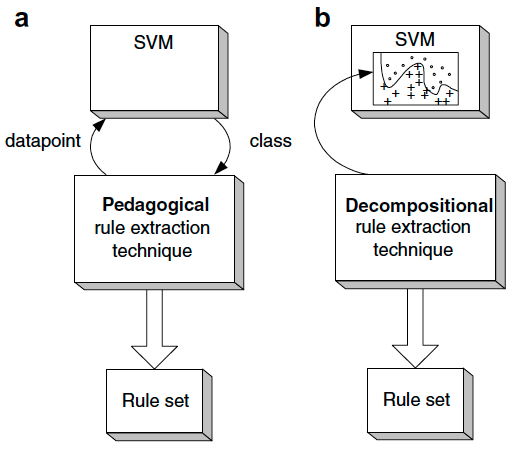}
	\caption{From \cite{martens2007comprehensible}: pedagogical \emph{(a)} and decompositional \emph{(b)} rule extraction techniques.}
	\label{fig:generalizable}
\end{figure}

\subsubsection{Explanation of Tree Ensembles.}
Finally, in \cite{deng2014interpreting}, Deng proposes the \emph{inTrees} framework $f$ to explain black boxes $b$ defined as \emph{Tree Ensembles (TE)} by returning a set of decision rules $c_g$.
In particular, InTrees extracts, measures, prunes and selects rules from tree ensembles, and calculates frequent variable interactions.
The set of black boxes $b$ that inTrees can explain is represented by any kind of tree ensemble like random forests, regularized random forests and boosted trees.
InTrees framework can be used for 
both classification and regression problems.
The technique described by InTrees is also known as \emph{Simplified Tree Ensamble Learner (STEL)}: it extracts the most supported and simplest rules form the trees ensemble. 

\subsection{Agnostic Explanator}
\label{sec:model_explanation_agnostic}
Recent approaches for interpretation are \emph{agnostic (AGN)} with respect to the black box to be explained.
In this section, we present a set of works solving the \emph{black box model explanation problem} by implementing function $f$ such that any type of black box $b$ can be explained.
These approaches do not return a specific comprehensible global predictor $c_g$, thus the type of explanation $\varepsilon_g$ change with respect to $f$ and $c_g$.
By definition all these approaches are generalizable.

Probably the first attempt of an agnostic solution was proposed in \cite{lou2012intelligible}.
Lou et al. propose a method $f$ which exploits Generalized Additive Models (GAMs) and it is able to interpret regression splines (linear and logistics), single trees and tree ensembles (bagged trees, boosted trees, boosted bagged trees and random forests).
GAMs are presented as the gold standard for intelligibility when only univariate terms are considered.
Indeed, the explanation $\varepsilon_c$ is returned as the importance of the contribution of the individual features in $b$ together with their \emph{shape function}, such that the impact of each predictor can be quantified.
A shape function is the plot of a function capturing the linearities and nonlinearities together with its shape.
It works on tabular data.
A refinement of the GAM approach is proposed by the same authors in \cite{lou2013accurate}.
A case study on health care showing the application of the GAM the refinement is presented in  \cite{caruana2015intelligible}. In particular, this approach is used for the prediction of the pneumonia risk and hospital 30-day readmission.

In \cite{henelius2014peek} the authors present an iterative algorithm $f$ that allows finding features and dependencies exploited by a classifier when producing a prediction. 
The attributes and the dependencies among the grouped attributes depict the global explanation $\varepsilon_g$.
The proposed approach $f$ named \emph{GoldenEye} is 
based on 
tabular data randomization (within class permutation, dataset permutation, etc.) and on  grouping attributes with interactions have an impact on the predictive power.

In \cite{krishnan2017palm}, \emph{PALM} is presented (\emph{Partition Aware Local Model}) to implement $f$.
In particular, PALM is a method that is able to learn and summarize the structure of the training dataset to help the machine learning debugging. 
PALM mimes a black box $b$ using 
a meta-model for partitioning the training dataset, and a set of sub-models for approximating and miming the patterns within each partition.
As meta-model it uses a decision tree ($c_g$) so that the user can examine its structure and determine if the rules detected follow the intuition or not, and link efficiently problematic test records to the responsible train data.
The sub-models linked to the leaves of the tree can be a arbitrarily complex model able to catch elaborate local patterns, but yet interpretable by humans. 
Thus, with respect to the final sub-models PALM is not only black box agnostic but also explanator agnostic.
Moreover, PALM is also data agnostic; i.e., it can work on any kind of data.

\subsection{Explanation via Other Approaches}
\label{sec:model_explanation_others}
In \cite{tolomei2017interpretable} a solution for the \emph{black box model explanation problem} is presented. It adopts an approach that can not be classified as one of the previous.
The proposed approach $f$ uses the internals of a random forest model $b$ to produce recommendations on the transformation of true negative examples into positively predicted examples.
These recommendations, which are strictly related to the feature importance, corresponds to the comprehensible global predictor $c_g$.
In particular, the function $f$ aims at transforming a negative instance into a positive instance by analyzing the path on the trees in the forest predicting such instance as positive or negative.
The explanation of $b$ is provided by means of the helpfulness of the features in the paths adopted for changing the instance outcome from negative to positive.

\section{Solving the Black Box Outcome Explanation Problem}
\label{sec:outcome_explanation}
In this section we review the methods solving the \emph{black box outcome explanation problem} (see Section~\ref{sec:outcome_explanation_problem}).
These methods provide a locally interpretable model which is able to explain the prediction of the black box in understandable terms for humans.
This category of approaches using a local point of view with respect to the prediction is becoming the most studied in the last years.
Section~\ref{sec:outcome_explanation_dnn} describes the methods providing the salient parts of the record for which a  prediction is required using \emph{Deep Neural Networks (DNNs)}, while Section~\ref{sec:outcome_explanation_agnostic} analyzes the methods which are able to provide a local explanation for any type of black box.

\subsection{Explanation of Deep Neural Network via Saliency Masks}
\label{sec:outcome_explanation_dnn}
In the following works the opened black box $b$ is a DNN and the explanation is provided by using a \emph{Saliency Mask (SM)}, i.e. a subset of the original record which is mainly responsible for the prediction.
For example, as salient mask we can consider the part of an image or a sentence in a text. A saliency image summarizes where a DNN looks into an image for recognizing their predictions.
The function $f$ to extract the local explanation $\varepsilon_l$ is always not generalizable and often strictly tied with the particular type of network, i.e., convolutional, recursive, etc. 

\begin{figure}[!tb]
\centering
	\includegraphics[trim = 0mm 0mm 0mm 0mm, clip, width=0.53\linewidth]{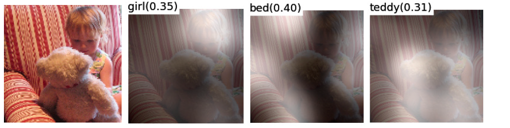}
	\includegraphics[trim = 0mm 0mm 0mm 0mm, clip, width=0.45\linewidth]{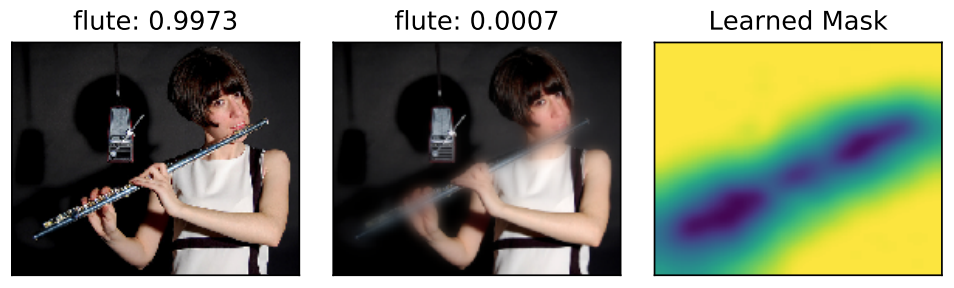}
	\caption{Saliency Masks for explanation of deep neural network. \emph{(Left)} From \cite{xu2015show} the elements of the image highlighted. \emph{(Right)} From \cite{fong2017interpretable} the mask and the level of accuracy on the image considering and not considering the learned mask.}
	\label{fig:dnn_examples}
\end{figure}

The work \cite{xu2015show} introduces an \emph{attention based model} $f$ which automatically identifies the contents of an image.
The black box is a neural network which consists of a combination of a \emph{Convolutional NN (CNN)} for the features extraction and a \emph{Recursive NN (RNN)} containing Long Short Term Memory (LSTM), nodes producing the image caption by generating a single word for each iteration.
The explanation $\varepsilon_l$ of the prediction is provided through a visualization of the attention (area of an image, see Figure~\ref{fig:dnn_examples}-\emph{left}) for each word in the caption.
A similar result is obtained by Fong et al. in \cite{fong2017interpretable}.
In this work the authors propose a framework $f$ of explanations $c_l$ as meta-predictors.
In their view, an explanation $\varepsilon_l$, and thus a meta-predictor, is a rule that predicts the response of a black box $b$ to certain inputs.
Moreover, they propose to use \emph{saliency maps} as explanations for black boxes to highlight the salient part of the images (see Figure~\ref{fig:dnn_examples}-\emph{right}).

Similarly, another set of works produce saliency masks incorporating network activations into their visualizations.
This kind of approaches $f$ are named \emph{Class Activation Mapping (CAM)}.
In \cite{zhou2016learning}, global average pooling in CNN (the black box $b$) is used for generating the CAM.
A CAM (the local explanation $\varepsilon_l$)  for a particular outcome label indicates the discriminative active region that identifies that label.
\cite{selvaraju2016grad} defines its relaxed generalization Grad-CAM which visualizes the linear combination of a late layer's activations and label-specific weights (or gradients for \cite{zhou2016learning}).
All these approaches arbitrarily invoke different back propagation and/or activation, which results in aesthetically pleasing, heuristic explanations of image saliency.
Their solution is not black box agnostic limited to NN, but it requires specific architectural modifications \cite{zhou2016learning} or access to intermediate layers \cite{selvaraju2016grad}.

With respect to texts, in \cite{lei2016rationalizing} the authors develop an approach $f$ which incorporates \emph{rationales} as part of the learning process of $b$.
A rationale is a simple subset of words representing a short and coherent piece of text (e.g., phrases), and alone must be sufficient for the prediction of the original text.
A rational is the local explanator $\varepsilon_l$ and provides the saliency of the text analyzed, i.e., indicates the reason for a certain outcome.

\subsection{Agnostic Explanator}
\label{sec:outcome_explanation_agnostic}
In this section we present the \emph{agnostic} solutions proposed for the \emph{black box outcome explanation problem} implementing function $f$ such that any type of black box $b$ can be explained.
All these approaches are generalizable by definition and return a comprehensible local predictor $c_l$.
Thus, they can be employed for diversified data types.

In \cite{ribeiro2016should}, Ribeiro et al. present the \emph{Local Interpretable Model-agnostic Explanations (LIME)} approach $f$ which does not depend on the type of data, nor on the type of black box $b$ to be opened, nor on a particular type of comprehensible local predictor $c_l$ or explanation $\varepsilon_l$.
In other words, LIME can return an understandable explanation for the prediction obtained by any black box.
The main intuition of LIME is that the explanation may be derived locally from the records generated randomly in the neighborhood of the record to be explained, and weighted according to their proximity to it.
In their experiments, the authors adopt linear models as comprehensible local predictor $c_l$ returning the importance of the features as explanation $\varepsilon_l$.
As black box $b$ the following classifiers are tested: decision trees, logistic regression, nearest neighbors, SVM and random forest. 
A weak point of this approach is the required transformation of any type of data in a binary format which is claimed to be human interpretable. 
\cite{ribeiro2016nothing} and \cite{ribeiro2016model} propose extensions of LIME with an analysis of particular aspects and cases.

A similar approach is presented in \cite{turner2016model}, where
Turner et al. design the \emph{Model Explanation System (MES)} $f$ that augments black box predictions with explanations by using a Monte Carlo algorithm.
In practice, they derive a scoring system for finding the best explanation based on formal requirements and consider that the explanations $\varepsilon_l$ are simple logical statements, i.e., decision rules. 
The authors test logistic regression and SVMs as black box $b$.


\section{Solving the Black Box Inspection Problem}
\label{sec:inspection}
In this section we review the methods for opening the black box facing the \emph{black box inspection problem} (see Section~\ref{sec:inspection_problem}).
Given a black box solving a classification problem, the inspection problem consists in providing a representation for understanding either how the black box model works or why the black box returns certain predictions more likely than others.
In \cite{seifert2017visualizations}, Seifet et al. provide a survey of visualizations of DNNs by defining a classification scheme describing visualization goals and methods. 
They found that most papers use pixel displays to show \emph{neuron activations}. 
As in the previous sections, in the following we propose a classification based on the type of technique $f$ used to provide the visual explanation of how the black box works.
Most papers in this section try to inspect NNs and DNNs.

\subsection{Inspection via Sensitivity Analysis}
\label{sec:inspection_sensitivity}
In this section we review the works solving the \emph{black box inspection problem} by implementing function $f$ using \emph{Sensitivity Analysis (SA)}.
Sensitivity analysis studies the correlation between the uncertainty in the output of a predictor and that one in its inputs \cite{saltelli2002sensitivity}.
All the following methods work on tabular datasets.

Sensitivity analysis for ``illuminating'' the black box was first proposed by Olden in  \cite{olden2002illuminating} where a visual method for understanding the mechanism of NN is described.
In particular, they propose to assess the importance of axon connections and the contribution of input variables by means of sensitivity analysis and \emph{Neural Interpretation Diagram (NID)} to remove not significant connections and improve the network interpretability.

In \cite{baehrens2010explain} the authors propose a procedure based on \emph{Gaussian Process Classification (GDP)} which allows explaining the decisions of any classification method through an explanation vector.
That is, the procedure $f$ is black box agnostic.
The explanation vectors are visualized to highlight the features that were most influential for the decision of a particular instance.
Thus, we are dealing with an inspection for outcome explanation $\varepsilon_l$. 

In \cite{datta2016algorithmic}, Datta et al. introduce a set of \emph{Quantitative Input Influence (QII)} measures $f$ capturing how much inputs influence the outputs of black box predictors. 
These measures provide a foundation for transparency reports of black box predictors.
In practice, the output consists in the feature importance for outcome predictions.

\cite{sundararajan2017axiomatic} studies the problem of attributing the prediction of a DNN (the black box $b$) to its input features. 
Two fundamental axioms are identified: sensitivity and implementation invariance. 
These axioms guide the design of an attribution method $f$, called \emph{Integrated Gradients (IG)}, that requires no modification to the original network.
Differently from the previous work, this approach is tested on different types of data. 

\begin{figure}[!tb]
\centering
	\includegraphics[trim = 0mm 0mm 0mm 0mm, clip, width=0.47\linewidth]{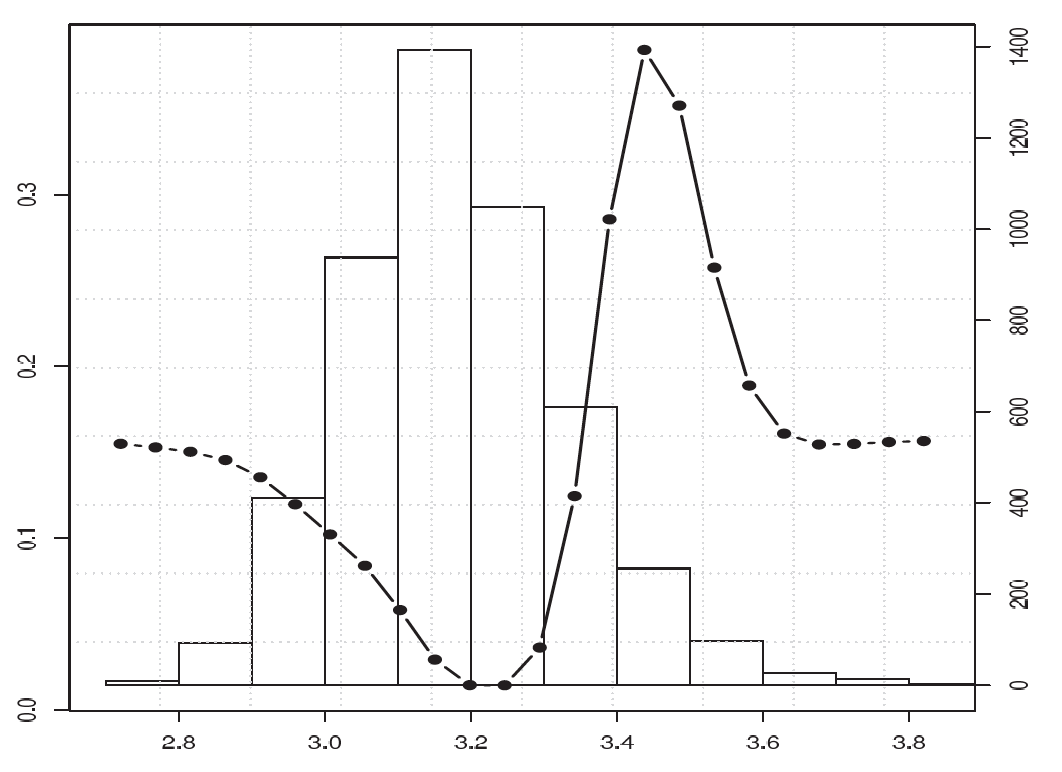}
	\hspace{5mm}
	\includegraphics[trim = 0mm 0mm 0mm 0mm, clip, width=0.47\linewidth]{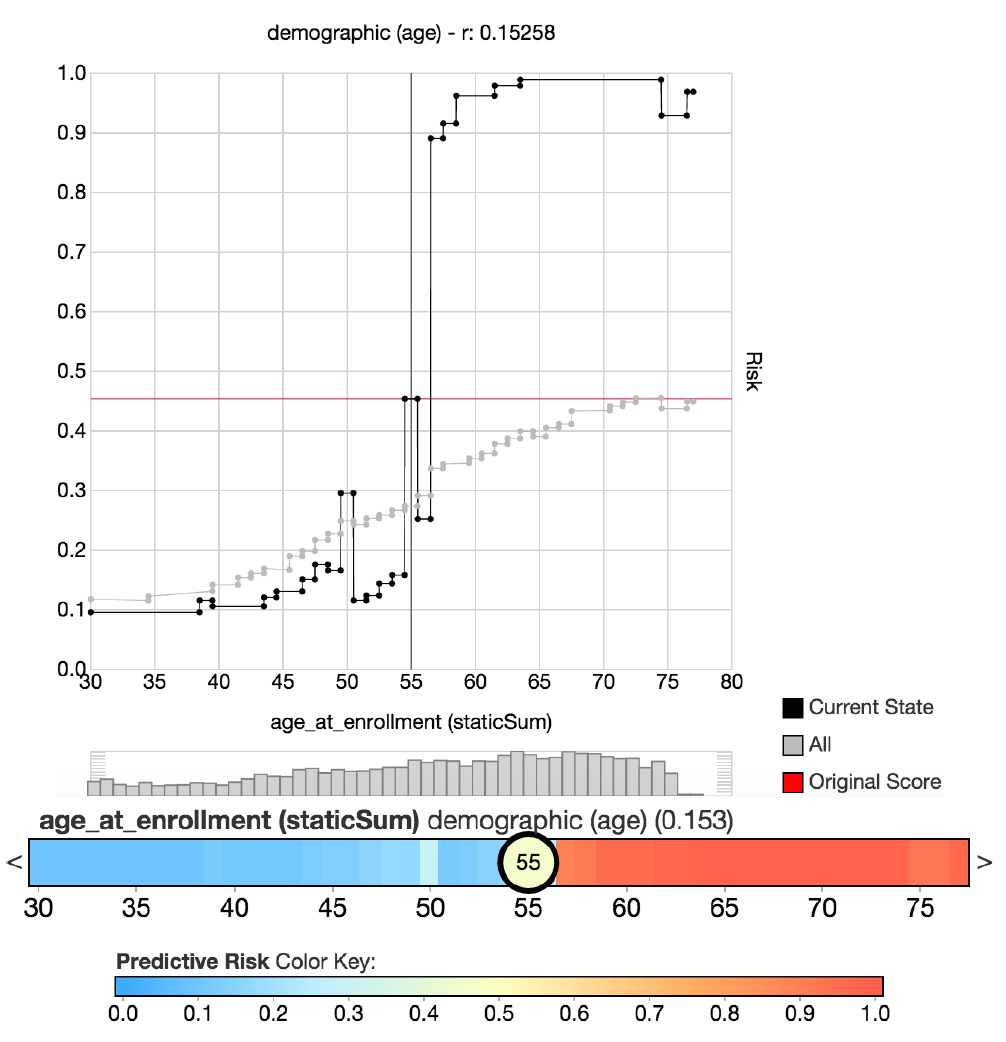}
		\caption{
		\emph{(Left)}. From \cite{cortez2011opening} VEC curve and histogram for the pH input feature (x-axis) and the respective high quality wine probability outcome (left of y-axis) and frequency (right of y-axis).
		\emph{(Right)}. From \cite{krause2016interacting} Age at enrollment shown as line plot (top) and partial dependence bar (middle). Color denotes the predicted risk of the outcome.
		}
	\label{fig:inspection_example}
\end{figure}

Finally, Cortez in \cite{cortez2011opening,cortez2013using} uses sensitivity analysis based and visualization techniques $f$ to explain black boxes $b$. 
The sensitivity measures are variables calculated as the range, gradient, variance of the prediction. 
Then, the visualizations realized are barplots for the features importance, and \emph{Variable Effect Characteristic curve (VEC)} \cite{cortez2009using} plotting the input values (x-axis) versus the (average) outcome responses (see Figure~\ref{fig:inspection_example} - \emph{(left)}).

\subsection{Inspection via Partial Dependence}
\label{sec:inspection_partial}
In this section we report a set of approaches solving the \emph{black box inspection problem} by implementing a function $f$ which returns a \emph{Partial Dependence Plot (PDP)}.
Partial dependence plot is a tool for visualizing the relationship between the response variable and predictor variables in a reduced feature space.
All the approaches presented in this section are black box agnostic and are tested on tabular datasets.

In \cite{hooker2004discovering}, the authors present an approach $f$ aimed at evaluating the importance of non-additive interactions between any set of features.
The implementation uses the \emph{Variable Interaction Network (VIN)} visualization generated from the use of ANOVA statistical methodology (a technique to calculate partial dependence plots).
VIN allows to visualize the importance of the features together with their interdependences.
Goldstein et al. provide in \cite{goldstein2015peeking} a technique $f$ which extends classical PDP named \emph{Individual Conditional Expectation (ICE)} to visualize the model approximated by a black box $b$ that help in visualizing the average partial relationship between the outcome and some features. 
ICE plots improves PDP by highlighting the variation in the fitted values across the range of covariates.
In \cite{krause2016interacting}, Krause et al. introduce random perturbations on the black box $b$ input values to understand to which extent every feature impact the prediction through a visual inspection using the PDPs $f$.
The main idea of \emph{Prospector} is to observe how the output varies by varying the input changing one variable at a time.
It provides an effective way to understand which are the most important features for a certain prediction $\varepsilon_l$ so that it can help in providing a valuable interpretation (see Figure \ref{fig:inspection_example} - \emph{(right)}). 
In \cite{adler2016auditing} the authors propose a method $f$ for \emph{auditing} (i.e., inspecting) black box predictors $b$, studying to which extent existing models benefit of specific features in the data. This method does not assume any knowledge on the models behavior.
In particular, the method $f$ focuses on \emph{indirect influence} and visualizes the global inspection $\varepsilon_g$ through an obscurity vs. accuracy plot (the features are obscured one after the other). 
Yet, the dependence of a black box $b$ on its input features is relatively quantified by the procedure $f$ proposed in \cite{adebayo2016iterative}, where 
the authors present an iterative procedure based on \emph{Orthogonal Projection of Input Attributes (OPIA)}, for enabling the interpretability of black box predictors.

\subsection{Inspection via Other Approaches}
\label{sec:inspection_others}
In the following, we present solutions for the \emph{black box inspection problem} that adopt an approach $f$ which can be categorized as none of the previous ones.
They all refer to DNNs as black box $b$ and are not generalizable.

\cite{yosinski2015understanding} proposes two tools for visualizing and interpreting DNNs and for understanding what computations DNNs perform at intermediate layers and which neurons activate.
These tools visualize the activations of each layer of a trained CNN during the process of images or videos.
Moreover, they visualize the features of the different layers by regularized optimization in image space.
Yosinski et al. found that by analyzing the live activations, changing in correspondence of different inputs, helps to generate an explanation on the DNNs behave. 
\cite{thiagarajan2016treeview} shows the extraction of a visual interpretation of a DNN using a decision tree. 
The method \emph{TreeView} $f$ works as follows.
Given the black box $b$ as a DNN, it first decomposes the feature space into K (user defined) potentially overlapping factors. 
Then, it builds a meta feature for each of the K clusters and a random forest that predicts the cluster labels. 
Finally, it generates and shows a surrogate decision tree from the forest as an approximation of the black box.

Shwartz-Ziv et al. in \cite{shwartz2017opening} showed the effectiveness of the \emph{Information Plane} $f$ visualization of DNNs highlighting that the empirical error minimization of each stochastic gradient descent phase epoch is always followed by a slow representation compression.

Finally, it is worth mentioning that \cite{radford2017learning} presents the discovery that a single neuron unit of a DNN can perform alone a sentiment analysis task after the training of the network reaching the same level of performance of strong baselines.
Also in \cite{zeiler2014visualizing}, Zeiler et al. backtrack the network computations to identify which image patches are responsible for certain neural activations. 

\section{Solving the Transparent Box Design Problem}
\label{sec:transparent_design}
In this section we review the approaches designed to solve the classification problem using a transparent method which is locally or globally interpretable on its own, i.e., solving the \emph{transparent box design problem} (see Section~\ref{sec:transparent_design_problem}).

\subsection{Explanation via Rule Extraction}
\label{sec:transparent_design_rule}
In this section we present the most relevant state of the art works solving the \emph{transparent box design problem} by means of comprehensible predictors $c$ based on \emph{rules}.
In these cases, $c_g$ is a comprehensible global predictor providing the whole set of rules leading to any possible decision: a \emph{global explanator} $\varepsilon_g$ is made available by $c_g$.
All the methods presented in this section work on tabular data.

In \cite{yin2003cpar} the authors propose the approach $f$ named \emph{CPAR (Classification based on Predictive Association Rules)} combining the positive aspects of both associative classification and traditional rule-based classification. 
Indeed, following the basic idea of FOIL \cite{quinlan1993foil}, CPAR does not generate a large set of candidates as in associative classification, and applies a greedy approach for generating rules $c_g$ directly from training data. 

Wang and Rudin, in \cite{wang2015falling} propose a method $f$ to extract falling rule lists $c_g$ (see Section \ref{sec:recognized_interpretable_models}) instead of classical rules.
The falling rule lists extraction method $f$ relies on a Bayesian framework.


In \cite{letham2015interpretable}, the authors tackle the problem to build a system for medical scoring which is interpretable and characterized by high accuracy. To this end, they propose \emph{Bayesian Rule Lists (BRL)} $f$ to extract the comprehensible global predictor $c_g$ as a \emph{decision list}. 
A decision list consists of a series of if-then statements discretizing the whole feature space into a set of simple and directly interpretable decision statements. 

A Bayesian approach is followed also in \cite{su2015interpretable}.
The authors propose algorithms $f$ for learning \emph{Two-Level Boolean Rules (TLBR)} in Conjunctive Normal Form or Disjunctive Normal Form $c_g$.
Two formulations are proposed. 
The first one is an integer program whose objective function combines the total number of errors and the total number of features used in the rule.
The second formulation replaces the 0-1 classification error with the Hamming distance from the current two-level rule to the closest rule that correctly classifies a sample. 
In \cite{lakkaraju2017interpretable} the authors propose a method $f$ exploiting a two-level boolean rule predictor to solve the black box model explanation, i.e., the transparent approach is used in the reverse engineering approach to explain the black box.

Yet another type of rule is exploited in \cite{lakkaraju2016interpretable}.
Here, Lakkaraju et al. propose a framework $f$ for generating prediction models, which are both interpretable and accuratem, by extracting \emph{Interpretable Decision Sets (IDS)} $c_g$, i.e., independent if-then rules. 
Since each rule is independently applicable, decision sets are simple, succinct, and easily to be interpreted. 
In particular, this approach can learn accurate, short, and non-overlapping rules covering the whole feature space. 

Rule Sets are adopted in \cite{wang2016bayesian} as comprehensible global predictor $c_g$.
The authors present a Bayesian framework $f$ for learning Rule Sets.
A set of parameters is provided to the user to encourage the model to have a desired size and shape in order to conform with a domain-specific definition of interpretability. 
A Rule Set consists of a small number of short rules where an instance is classified as positive if it satisfies at least one of the rules. 
The rule set provides reasons for predictions, and also descriptions of a particular class.

Finally, in \cite{malioutov2017learning} an approach $f$ is designed to learn both
sparse \textit{conjunctive} and \textit{disjunctive} clause rules from training data through a linear programming solution. The optimization formulation leads the resulting rule-based global predictor $c_g$ (\emph{1Rule}) to automatically balance accuracy and interpretability.

\subsection{Explanation via Prototype Selection}
\label{sec:transparent_design_prototype}
In this section we present the design of a set of approaches $f$ for solving the \emph{transparent box design problem} returning a comprehensible predictor $c_g$ equipped with a human understandable global explanator function $\varepsilon_g$.
A prototype, also referred to with the name artifact or archetype, is an object that is representative of a set of similar instances.
A prototype can be an instance $x$ part of the training set $D = \{ X, Y \}$, or can lie anywhere in the space $\mathcal{X}^m \times \mathcal{Y}$ of the dataset $D$.
Having only prototypes among the observed points is desirable for interpretability, but it can also improve the classification error.
As an example of a prototype we can consider the record minimizing the sum of the distances with all the other points of a set (like in K-Medoids) or the record generated averaging the value of the features of a set of points (like in K-Means) \cite{tan2006introduction}.
Different definitions and requirements to find a prototype are specified in each work using the prototypes to explain the black box.

In \cite{bien2011prototype}, Bien et al. design the transparent \emph{Prototype Selection (PS)} approach $f$ that first seeks for the best prototype (two strategies are proposed), and then assigns the points in $D$ to the label corresponding to the prototype.
In particular, they face the problem of recognizing hand written digits.
In this approach, every instance can be described by more than one prototype, and more than a prototype can refer to the same label (e.g., there can be more than one prototype for digit zero, more than one for digit one, etc.).
The comprehensible predictor $c_g$ provides a global explanation in which every instance must have a prototype corresponding to its label in its neighborhood; no instances should have a prototype with a different label in its neighborhood, and there should be as few prototypes as possible.

Kim et al. in \cite{kim2014bayesian,kim2015mind} design the \emph{Bayesian Case Model (BCM)} comprehensible predictor $c_l$ able to learn prototypes by clustering the data and to learn subspaces. Each prototype is the representative sample of a given cluster, while the subspaces are set of features which are important in identifying the cluster prototype. 
That is, the global explanator $\varepsilon_g$ returns a set of prototypes together with their fundamental features.
Possible drawbacks of this approach are the high number of parameters (e.g., number of clusters) and various types of probability distributions which are assumed to be correct for each type of data.
\cite{kim2015ibcm} proposes an extension of BCM which exploits humans interaction to improve the prototypes. 
Finally, in \cite{kim2016examples} the approach is further expanded to include criticisms, where a criticism is an instance that does not fit the model very well, i.e., a counter-example part of the cluster of a prototype.

With respect to prototypes and DNN, \cite{mahendran2015understanding} proposes a method $b$ to change the image representations in order to use only information from the original image representation and from a generic natural image prior. 
This task is mainly related to image reconstruction rather than black box explanation, but it is realized with the aim of understanding the example to which the DNN $b$ is related to producing a certain  prediction by realizing a sort of artificial image prototype.
There is a significant amount of work in understanding the representation of DNN by means of artifact images, \cite{kato2014image,vondrick2013hoggles,weinzaepfel2011reconstructing,zeiler2014visualizing}. 

We conclude this section presenting how \cite{fong2017interpretable} deals with artifacts in DNNs.
Finding a single representative prototype by perturbation, deletion, preservation, and similar approaches has the risk of triggering artifacts of the black box.
As discussed in Section~\ref{sec:inspection_others}, NN and DNN are known to be affected by surprising artifacts. 
For example, \cite{kurakin2016adversarial} shows that a nearly-invisible image perturbation can lead a NN to classify an object for another; \cite{nguyen2015deep} constructs abstract synthetic images that are classified arbitrarily; \cite{mahendran2015understanding} finds deconstructed versions of an image which are indistinguishable from the viewpoint of the DNN from the original image, and also with respect to texts \cite{liang2017deep} inserts typos and random sentences in real texts that are classified arbitrarily.
These examples demonstrate that it is possible to find particular inputs that can drive the DNN to generate nonsensical or unexpected outputs. 
While not all artifacts look ``unnatural'', nevertheless they form a subset of images that are sampled with negligible probability when the network is normally operated.
In our opinion, two guidelines should be followed to avoid such artifacts in generating explanations for DNNs, and for every black box in general. 
The first one is that powerful explanations should, just like any predictor, generalize as much as possible. 
Second, the artifacts should not be representative of natural perturbations.

\subsection{Explanation via Other Approaches}
\label{sec:transparent_design_others}
In the following we present solutions for the \emph{transparent box design problem} adopting approaches $f$ that can not be categorized as the previous ones.
\cite{kononenko2010efficient} describes a method $f$ based on Naive Bayes aimed to explain individual predictions $\varepsilon_l$ of black boxes $b$.
The proposed approach exploits notions from \emph{coalitional game theory}, and explains predictions utilizing the contribution of the value of different individual features $\varepsilon_l$ (see Figure~\ref{fig:linearmodel}).
The method is agnostic with respect to the black box used and is tested only on tabular data.
Finally, in \cite{wang2015trading} Wang et al. propose a method $f$ named \emph{OT-SpAMs} based on oblique tree sparse additive models for obtaining a global interpretable predictor $c_g$ as a decision tree.
\emph{OT-SpAMs} divides the feature space into regions using a sparse oblique tree splitting and assigns local sparse additive experts (leaf of the tree) to individual regions.
Basically, \emph{OT-SpAMs} passes from complicated trees/linear models to an explainable tree $\varepsilon_g$.

\section{Conclusion}
\label{sec:conclusion}


In this paper we have presented a comprehensive overview of methods proposed in the literature for explaining decision systems based on opaque and obscure machine learning models.
First, we have identified the different components of the family of the explanation problems. 
In particular, we have provided a formal definition of each problem belonging to that family capturing for each one the proper peculiarity. 
We have named these problems: \textit{black box model explanation problem}, \textit{black box outcome explanation problem}, \textit{black box inspection problem} and \textit{transparent box design problem}.
Then, we have proposed a classification of methods studied in the literature which take into account the following dimensions: the specific explanation problem addressed, the type of explanator adopted, the black box model opened, and the type of data used as input by the black box model. 

As shown in this paper, a considerable amount of work has already been done in different scientific communities and especially in the machine learning and data mining communities. 
The first one is mostly focused on describing how the black boxes work, while the second one is more interested into explaining the decisions even without understanding the details on how the opaque decision systems work in general.

The analysis of the literature conducted in this paper has led to the conclusion that despite many approaches have been proposed to explain black boxes, some important scientific questions still remain unanswered. 
One of the most important open problems is that, until now, there is no agreement on what an \textit{explanation} is. 
Indeed, some works provide as explanation a set of rules, others a decision tree, others a prototype (especially in the context of images). 
It is evident that the research activity in this field completely ignored the importance of studying a general and common formalism for defining an explanation, identifying which are the \textit{properties} that an explanation should guarantee, e.g., soundness, completeness, compactness and comprehensibility.
Concerning this last property, there is no work that seriously addresses the problem of quantifying the grade of comprehensibility of an explanation for humans, although it is of fundamental importance.
The study of measures able to capture this aspect is challenging because it also consider also aspects like the expertise of the user or the amount of time available to understand the explanation.
The definition of a (mathematical) formalism for explanations and of tools for measuring how much an explanation is comprehensible for humans would improve the practical applicability of most of the approaches presented in this paper. 

Moreover, there are other open research questions related to black boxes and explanations that are starting to be treated by the scientific community and that deserve attention and more investigation.

A common assumption of all categories of works presented in this paper is that the features used by the black box decision system are completely known. 
However, a black box might use additional information besides that explicitly asked to the user. 
For example, it might link the user's information with different data sources for augmenting the data to be exploited for the prediction.
Therefore, an important aspect to be investigated is to understand how an explanation might also be derived in cases where black box systems make decisions in presence of \emph{latent features}. 
An interesting starting point for this research direction is the framework proposed in \cite{lakkaraju2017selective} by Lakkaraju et al. for the evaluation of the prediction models performances on 
labeled data where the decision of decision-makers (either humans or black-boxes) is taken in the presence of unobserved features.

Another open research question is related to providing explanations in the field of \emph{recommender systems}.
When a suggestion is provided to a customer, it should come together with the reasons for this recommendation. 
In \cite{mcsherry2005explanation} the authors define a case-based reasoning approach to generate recommendations with the opportunity of obtaining both the explanation of the recommendation process and of the produced recommendations. 

Lastly, a further interesting point is the fact that explanations are important on their own and predictors might be learned directly from explanations.
A starting study of this aspect is \cite{krening2017learning} that presents a software agent learned to simulate the Mario Bros game only utilizing explanations rather than the logs of previous plays.

\appendix

\section{Supplementary Materials}
\label{sec:supplementary_material}


\newcolumntype{L}[1]{>{\raggedright\let\newline\\\arraybackslash\hspace{0pt}}m{#1}}

\begin{table}[b]
    \centering
    \setlength{\tabcolsep}{2mm}
\caption{Summary of methods for opening and explaining black boxes with respect to the problem faced.}
    \begin{tabular}{cL{10cm}}
    \hline
    \textbf{Problem} & \textbf{References} \\
    \hline
    \rowcolor{gray!15} \emph{Model Explanation} & 
        \cite{craven1996extracting}, 
        \cite{krishnan1999extracting},
        \cite{boz2002extracting},
        \cite{johansson2009evolving},
        \cite{chipman1998making},
        \cite{gibbons2013cad},
        \cite{zhou2016interpreting},
        \cite{schetinin2007confident},
        \cite{hara2016making},
        \cite{tan2016tree},
        \cite{andrews1995survey},
        \cite{craven1994using},
        \cite{johansson2003rule},
        \cite{zhou2003extracting},
        \cite{augasta2012reverse},
        \cite{nunez2002rule},
        \cite{fung2005rule},
        \cite{martens2007comprehensible},
        \cite{deng2014interpreting},
        \cite{lou2013accurate},
        \cite{henelius2014peek},
        \cite{krishnan2017palm},
        \cite{tolomei2017interpretable}
    \\
    \emph{Outcome Explanation} & 
        \cite{xu2015show},
        \cite{fong2017interpretable},
        \cite{zhou2016learning},
        \cite{selvaraju2016grad},
        \cite{lei2016rationalizing},
        \cite{ribeiro2016nothing},
        \cite{turner2016model}
    \\
    \rowcolor{gray!15} \emph{Black Box Inspection} & 
        \cite{olden2002illuminating},
        \cite{baehrens2010explain},
        \cite{sundararajan2017axiomatic},
        \cite{cortez2011opening},
        \cite{hooker2004discovering},
        \cite{goldstein2015peeking},
        \cite{krause2016interacting},
        \cite{adler2016auditing},
        \cite{adebayo2016iterative},
        \cite{yosinski2015understanding},
        \cite{thiagarajan2016treeview},
        \cite{shwartz2017opening},
        \cite{radford2017learning}
    \\
    \emph{Transparent Design} & 
        \cite{yin2003cpar},
        \cite{wang2015falling},
        \cite{letham2015interpretable},
        \cite{su2015interpretable},
        \cite{lakkaraju2016interpretable},
        \cite{wang2016bayesian},
        \cite{malioutov2017learning},
        \cite{bien2011prototype},
        \cite{kim2014bayesian},
        \cite{mahendran2015understanding},
        \cite{kononenko2010efficient},
        \cite{wang2015trading}
    \\
    \hline
    \end{tabular}
    \label{tab:problem}
\end{table}

\begin{table}[b]
    \centering
    \setlength{\tabcolsep}{2mm}
\caption{Summary of methods for opening and explaining black boxes with respect to the explanator adopted.}
    \begin{tabular}{cL{9cm}}
    \hline
    \textbf{Explanator} & \textbf{References} \\
    \hline
    \rowcolor{gray!15} \emph{Decition Tree (DT)} & 
        \cite{craven1996extracting}, 
        \cite{krishnan1999extracting},
        \cite{boz2002extracting},
        \cite{johansson2009evolving},
        \cite{chipman1998making},
        \cite{domingos1998knowledge},
        \cite{gibbons2013cad},
        \cite{zhou2016interpreting},
        \mbox{\cite{schetinin2007confident},
        \cite{hara2016making},
        \cite{tan2016tree},
        \cite{krishnan2017palm},
        \cite{thiagarajan2016treeview},
        \cite{wang2015trading}}
    \\
    \emph{Decision Rules (DR)} & 
        \cite{andrews1995survey},
        \cite{craven1994using},
        \cite{johansson2003rule},
        \cite{zhou2003extracting},
        \cite{augasta2012reverse},
        \cite{nunez2002rule},
        \cite{fung2005rule},
        \cite{martens2007comprehensible},
        \cite{deng2014interpreting},
        \mbox{\cite{turner2016model},
        \cite{yin2003cpar},
        \cite{wang2015falling},
        \cite{letham2015interpretable},
        \cite{su2015interpretable},
        \cite{lakkaraju2016interpretable},
        \cite{wang2016bayesian},
        \cite{malioutov2017learning}}
    \\
    \rowcolor{gray!15} \emph{Features Importance (FI)} & 
        \cite{lou2013accurate},
        \cite{henelius2014peek},
        \cite{tolomei2017interpretable},
        \cite{ribeiro2016nothing}
    \\
    \emph{Saliency Mask (SM)} & 
        \cite{xu2015show},
        \cite{fong2017interpretable},
        \cite{zhou2016learning},
        \cite{selvaraju2016grad},
        \cite{lei2016rationalizing}
    \\
    \rowcolor{gray!15} \emph{Sensitivity Analysis (SA)} & 
        \cite{olden2002illuminating},
        \cite{baehrens2010explain},
        \cite{sundararajan2017axiomatic},
        \cite{cortez2011opening}
    \\
    \emph{Partial Dependence Plot (PDP)} & 
        \cite{hooker2004discovering},
        \cite{goldstein2015peeking},
        \cite{krause2016interacting},
        \cite{adler2016auditing},
        \cite{adebayo2016iterative}
    \\
    \rowcolor{gray!15} \emph{Neurons Activation (NA)} & 
        \cite{yosinski2015understanding},
        \cite{shwartz2017opening},
        \cite{radford2017learning}
    \\
    \emph{Prototype Selection (PS)} & 
        \cite{bien2011prototype},
        \cite{kim2014bayesian},
        \cite{mahendran2015understanding}
    \\
    \hline
    \end{tabular}
    \label{tab:explanator}
\end{table}

\begin{table}[b]
    \centering
    \setlength{\tabcolsep}{2mm}
\caption{Summary of methods for opening and explaining black boxes with respect to the black box explained.}
    \begin{tabular}{cL{9cm}}
    \hline
    \textbf{Black Box} & \textbf{References} \\
    \hline
    \rowcolor{gray!15} \emph{Neural Network (NN)} &
        \cite{craven1996extracting}, 
        \cite{krishnan1999extracting},
        \cite{boz2002extracting},
        \cite{johansson2009evolving},
        \cite{domingos1998knowledge},
        \cite{andrews1995survey},
        \cite{craven1994using},
        \cite{johansson2003rule},
        \cite{zhou2003extracting},
        \cite{augasta2012reverse},
        \cite{olden2002illuminating}
    \\
    \emph{Tree Ensemble (TE)} & 
        \cite{chipman1998making},
        \cite{domingos1998knowledge},
        \cite{gibbons2013cad},
        \cite{zhou2016interpreting},
        \cite{schetinin2007confident},
        \cite{hara2016making},
        \cite{tan2016tree},
        \cite{deng2014interpreting},
        \cite{tolomei2017interpretable}
    \\
    \rowcolor{gray!15} \emph{Support Vector Machines (SVM)} & 
        \cite{nunez2002rule},
        \cite{fung2005rule},
        \cite{martens2007comprehensible}
    \\
    \emph{Deep Neural Network (DNN)} & 
        \cite{xu2015show},
        \cite{fong2017interpretable},
        \cite{zhou2016learning},
        \cite{selvaraju2016grad},
        \cite{lei2016rationalizing},
        \cite{sundararajan2017axiomatic},
        \cite{yosinski2015understanding},
        \cite{thiagarajan2016treeview},
        \cite{shwartz2017opening},
        \cite{radford2017learning}
    \\
    \rowcolor{gray!15} \emph{Agnostic black box (AGN)} & 
        \cite{lou2013accurate},
        \cite{henelius2014peek},
        \cite{krishnan2017palm},
        \cite{ribeiro2016nothing},
        \cite{turner2016model},
        \cite{baehrens2010explain},
        \cite{cortez2011opening},
        \cite{hooker2004discovering},
        \cite{goldstein2015peeking},
        \cite{krause2016interacting},
        \cite{adler2016auditing},
        \cite{adebayo2016iterative}
    \\
    \hline
    \end{tabular}
    \label{tab:blackbox}
\end{table}


\section*{Acknowledgement}
This work is partially supported by the European Community's H2020 Program under the funding scheme ``INFRAIA-1-2014-2015: Research Infrastructures'', grant agreement 654024, \emph{SoBigData}, \url{http://www.sobigdata.eu}.

\bibliographystyle{abbrv}
\bibliography{biblio}

\end{document}